\documentclass [11pt,letterpaper]{JHEP3}
\usepackage{amsmath, amsfonts}
\usepackage{yfonts}
\usepackage{epsfig}
\usepackage{cite}	
\usepackage{bm}		
\usepackage{verbatim}   
\usepackage{psfrag}     

\advance\parskip 2.0pt plus 1.0pt minus 2.0pt	

\let\hreff=\href
\def\href#1#2{#2}	

\def\n{{\bf n}}
\def\m{{\bf m}}
\def\C{{\cal C}}

\def\O{{\cal O}}
\def\Otilde{\widetilde \O}
\def\tr{{\rm tr}}

\def\Nc{N_{\rm c}}
\def\Nf{N_{\rm f}}
\def\nf{n_{\rm f}}

\def\Z{{\mathbb Z}}

\def\Dslash{{\rlap{\raise 1pt \hbox{$\>/$}}D}}
\def\None{\mathcal N\,{=}\,1}

\def\R{{\bf R}}

\preprint {
	NSF-KITP-07-07\\
	SLAC-PUB-12331}

\title
    {%
    Volume independence in large $\bm \Nc$ QCD-like gauge theories
    }

\author
    {%
    Pavel~Kovtun,\!$^1$\footnote{\email{kovtun@kitp.ucsb.edu}}~
    Mithat \"Unsal,\!$^{2}$\footnote{\email{unsal@slac.stanford.edu}}~
    and Laurence~G.~Yaffe$^3$\footnote{\email{yaffe@phys.washington.edu}}
    \\$^1$KITP, University of California, Santa Barbara, CA 93106--4030
    \\$^2$SLAC and Department of Physics, Stanford University, CA 94025
    \\$^3$Department of Physics, University of Washington, Seattle,
	  WA 98195--1560
    \\
    }%

\abstract
   {%
   Volume independence in large $\Nc$ gauge theories
   may be  viewed as a generalized orbifold equivalence.
   The reduction to zero volume (or Eguchi-Kawai reduction)
   is a special case of this equivalence.
   So is temperature independence in confining phases.
   A natural generalization concerns
   volume independence in ``theory space'' of quiver gauge theories.
   In pure Yang-Mills theory,
   the failure of volume independence for sufficiently small volumes
   (at weak coupling)
   due to spontaneous breaking of center symmetry,
   together with its validity above a critical size,
   nicely illustrate the symmetry realization conditions which are
   both necessary and sufficient for large $\Nc$ orbifold equivalence.
   The existence of a minimal size below which volume independence fails
   also applies to Yang-Mills theory with
   antisymmetric representation fermions [QCD(AS)].
   However, in Yang-Mills theory with adjoint representation fermions
   [QCD(Adj)], endowed with periodic boundary conditions,
   volume independence remains valid down to arbitrarily small size.
   In sufficiently large volumes,
   QCD(Adj) and QCD(AS) have a large $\Nc$ ``orientifold'' equivalence,
   provided charge conjugation symmetry is unbroken in the latter theory.
   Therefore,
   via a combined orbifold-orientifold mapping,
   a well-defined large $\Nc$ equivalence exists between
   QCD(AS) in large, or infinite, volume and QCD(Adj) in arbitrarily small
   volume.
   Since asymptotically free gauge theories, such as QCD(Adj),
   are much easier to study (analytically or numerically)
   in small volume, this equivalence should allow greater understanding of
   large $\Nc$ QCD in infinite volume.
   }%


\begin {document}

\setlength{\baselineskip}{1.10\baselineskip}
\let\href=\hreff

\section {Introduction and Summary}

Back in 1982,
Eguchi and Kawai \cite{Eguchi-Kawai} argued
that properties of $U(\Nc)$ Yang-Mill theory,
formulated on a periodic lattice,
are independent of the lattice size in the $\Nc\to\infty$ limit.
It was quickly understood that this large $\Nc$ equivalence
(relating a pair of theories in different volumes)
is valid only when certain global symmetries are not spontaneously
broken \cite{LGY-largeN,BHN},
and applies only to appropriate observables.
When the theory is defined, for simplicity,
in a $d$-dimensional periodic hypercube
of size $L$, recent work
has shown that large $\Nc$ volume independence is valid (in $d > 2$)
only above a critical size, $L > L_c(d)$.
The critical size $L_c(d)$ depends on the lattice spacing,
in addition to dimension,
but approaches a finite physical value in the continuum limit
\cite{Narayanan-Neuberger,Kiskis-Narayanan-Neuberger}.

Large-$\Nc$ equivalences between pairs of
theories related by so-called ``orbifold'' projections
have also received considerable attention in recent years.
(See, for example, Refs.~%
\cite{Bershadsky-Johansen,Tong,KUY1,ASV1,ASV2,KUY2,KUY3,Barbon}.)
In this context, orbifold projection is a technique for constructing
a ``daughter'' theory, starting from some ``parent'' theory,
by retaining only those fields which are invariant under a
discrete symmetry group of the parent theory.
For orbifold projections based on cyclic groups,
Refs.~\cite{KUY1,KUY2} proved
necessary and sufficient conditions for
equivalence of parent and daughter theories in the large $\Nc$
limit, and clarified previous confusion in the literature
regarding the appropriate mapping of information between theories.
In complete analogy with volume independence,
large $\Nc$ orbifold equivalence is applicable only
when certain global symmetries are not spontaneously broken,
and applies only to a restricted class of observables.

In this paper, we discuss the relation between
volume independence and non-perturbative orbifold equivalence
in the large $\Nc$ limit.
The mapping which takes a $U(\Nc)$ gauge theory in
volume $L^d$ to the same theory in a smaller volume $(L')^d$,
with $L/L'$ any integer,
may be viewed as a generalized orbifold projection
which eliminates degrees of freedom that are not invariant
under a discrete translation group.%
\footnote
    {
    In this paper, we are only concerned with theories formulated
    on toriodal compactifications of flat space.
    }
We show that inverse mappings, which blow up a smaller volume theory
into a larger volume, may also be understood as orbifold projections,
specifically ones which eliminate degrees of freedom that are not
invariant under a discrete Abelian subgroup of the gauge (and flavor)
symmetry group.  Interpreting these volume-changing mappings as orbifold
projections applies not only to pure Yang-Mills theories, but also to
theories with fundamental, adjoint, or rank-two tensor representation
matter fields.%
\footnote
    {
    For earlier, somewhat related work, see
    Refs.~\cite{Orland,Neuberger}.
    }

Viewing volume changing mappings between gauge theories as
generalized orbifold projections allows a simple proof of large
$\Nc$ equivalence using the methods of Refs.~\cite{KUY1,KUY2},
and also clarifies how the large $\Nc$ equivalence applies to
connected correlators (in addition to expectation values)
of operators in the appropriate ``neutral'' sectors
of either theory.
The validity of large $\Nc$ orbifold equivalence depends on
certain symmetry realization conditions --- the discrete symmetries used
in either direction of these projections must not be spontaneously broken.
These symmetry realization conditions are both necessary and sufficient
\cite{KUY2}.
As we discuss, the numerical results of Ref.~\cite{Kiskis-Narayanan-Neuberger}
are completely in accord with, and nicely illustrate, these general
conditions.

A special case of large $\Nc$ volume independence, which we discuss,
is large $\Nc$ temperature independence.
For mappings which change the size of a single periodic dimension,
interpreted as the Euclidean time direction,
the symmetry realization condition necessary for large $\Nc$ equivalence
is just the condition that the theory be in a confining phase.
Large $\Nc$ temperature independence implies that the leading large $\Nc$
behavior of expectation values and connected correlators of
suitable operators are temperature independent in any confining phase.%
\footnote
    {
    For previous discussion of the confinement/deconfinement transition
    in the context of reduced models, see Ref.~\cite{Neuberger:1983xc}.
    }

In contrast to other examples of large $\Nc$ orbifold (or ``orientifold'')
equivalences, such as those discussed in
Refs.~\cite{Bershadsky-Johansen,Tong,ASV1,ASV2},
large $\Nc$ volume (or temperature) independence cannot
be seen in a perturbative analysis --- it is intrinsically non-perturbative.
As we will discuss, this reflects the fact that unbroken center symmetry
is a necessary condition for large $\Nc$ volume independence, and
perturbative expansions do not respect the center symmetry.

For later convenience, let QCD(Adj) denote a $U(\Nc)$ gauge
theory with one or more massless adjoint representation Majorana fermions,
and let QCD(AS) [or QCD(S)] denote a $U(\Nc)$ gauge theory
with one or more massless Dirac fermions in the rank-two
antisymmetric [or symmetric] tensor representation.%
\footnote
    {
    We will also use QCD(AS/S) to denote either QCD(AS) or QCD(S)
    in contexts where the same results apply to either theory.
    }
Single-flavor massless QCD(Adj) is precisely
$\None$ supersymmetric Yang-Mills theory.
And at $\Nc = 3$, QCD(AS) is the same
as ordinary QCD with fundamental representation fermions.
Hence, QCD(AS) is a generalization of ordinary QCD to arbitrary $\Nc$
in which a single flavor of fermions will continue to play
a significant role even at $\Nc = \infty$ \cite{ASV1,ASV2}.

In pure $U(\Nc)$ Yang-Mills theory,
as noted above,
volume independence fails for sufficiently small volumes
due to a change in symmetry realization.
But, as we discuss,
the addition of massless adjoint representation fermions
(with periodic boundary conditions)
prevents this phase transition.
Consequently volume independence
in QCD(Adj) remains valid down to arbitrarily small volumes.

This volume independence of QCD(Adj) has interesting implications
for large $\Nc$ QCD,
because a separate large-$\Nc$ equivalence \cite{ASV1,Armoni:2004ub}
relates QCD(Adj) and QCD(AS),
provided the latter theory does not spontaneously break
charge conjugation symmetry \cite{UY}.
In sufficiently small volumes
(with periodic boundary conditions for the fermions)
QCD(AS) {\em does\/} undergo charge conjugation symmetry breaking,
thus invalidating the large-$\Nc$ equivalence to QCD(Adj) \cite{UY}.
However, all available evidence is consistent with
QCD(AS) having unbroken charge conjugation symmetry in
sufficiently large volume \cite{DeGrand:2006qb}.
Provided this is true,
combining the large-$\Nc$ equivalence between QCD(AS)
and QCD(Adj), in sufficiently large volume,
with the large-$\Nc$ volume independence of QCD(Adj),
allows one to relate properties of QCD(AS) in large (or infinite)
volume to corresponding quantities in QCD(Adj)
in arbitrarily small volume.
As we will discuss,
this equivalence applies to both expectation values and
connected correlation functions of suitable (charge conjugation even)
bosonic operators.
In the zero volume limit, this shows that a simple matrix model
(or ``reduced'' theory) can reproduce the leading
large $\Nc$ behavior of a large class of correlation functions in
infinite volume QCD(AS) --- which is a natural generalization
of real QCD to large $\Nc$.%
\footnote
    {
    This formulation of a reduced matrix model for QCD,
    in which fermions prevent the spontaneous breaking of
    center symmetry,
    is complementary to previous approaches yielding
    ``quenched'' \cite{BHN}
    or ``twisted'' \cite{Gonzalez-Arroyo:1982ub,Gonzalez-Arroyo:1982hz}
    reduced models,
    where modifications to the action are introduced by hand
    in order to prevent unwanted symmetry breaking.
    See Ref.~\cite{Makeenko} for a good pedagogical discussion
    of quenched and twisted reduced models.
    In particular,
    these approaches lack direct applicability to connected correlators
    (but see, however, Ref.~\cite{Levine:1982fa}).
    }

Following this extended discussion of volume independence in spacetime,
we show that a completely analogous approach demonstrates
``theory space volume independence'' in quiver gauge theories.
These are theories with product gauge groups and bifundamental
matter fields.
``Theory space'' is a convenient graphical representation of
the gauge group and matter field content of such theories
\cite{Rothstein:2001tu, Arkani-Hamed:2001ed}.
We show that one may reduce a quiver gauge theory
with multiple equivalent gauge group factors
to another quiver theory
with a smaller product gauge group by performing a projection
based on a subgroup of the symmetry group permuting equivalent
gauge group factors.
Alternatively, a quiver gauge theory can be ``blown up''
to produce bigger quiver theories with enlarged product gauge groups
by performing orbifold projections based on a subgroup involving
the global flavor symmetry.
Projections which increase the volume
(in either spacetime or theory space) act as inverses
of the projections which reduce volume.%
\footnote
    {
    This, of course, is somewhat sloppy terminology
    since both projections reduce the number of degrees of freedom
    in a theory.
    The net effect of a projection times its ``inverse'' is
    a reduction of $\Nc$ by a fixed multiplicative factor,
    while leaving the form of the theory otherwise unchanged.
    }
In all cases, the condition that the relevant discrete symmetries
used in the projections
not be spontaneously broken is both necessary and sufficient for
large-$\Nc$ equivalence between pairs of quiver gauge theories
related by these orbifold projections.

It should be noted that generic orbifold projections
do not produce examples of large $\Nc$ equivalence
(even when one requires that neither the parent nor daughter theories
break any symmetries spontaneously).%
\footnote
    {
    For example, a $\Z_2$ orbifold projection can take a
    $U((k+l)N)$ theory with adjoint matter to a
    $U(kN) \times U(lN)$ theory with bifundamental matter.
    But, unless $k = l$, there is no simple mapping which relates
    expectation values or correlators of
    observables of these theories in the $N\to\infty$ limit.
    }
Precise conditions which delineate which generalized orbifold projections
do, or do not, lead to examples of large $\Nc$ equivalence
are not currently known.
However, our results support a simple and natural conjecture:
\goodbreak

\begin{quote}
    Two theories, $A$ and $B$, each with smooth large $\Nc$ limits,
    will have coinciding large $\Nc$ limits,
    within appropriate neutral sectors,
    if a generalized orbifold projection exists which
    maps theory $A$ with $\Nc = k N$ to theory $B$ with $\Nc = N$,
    for some $k$,
    and an ``inverse'' orbifold projection exists which maps theory $B$
    with $\Nc = m N$ back to theory $A$ with $\Nc = N$, for some $m$.
\end{quote}
In addition to the examples discussed in this paper,
and their obvious generalizations, this conjecture is
also supported by a wide variety of examples involving
mappings between unitary, orthogonal, and symplectic gauge groups.
This is discussed in a forthcoming paper \cite{KUY5}.

\section{Spacetime volume independence}
\label{sec:EK}

\subsection{Volume reduction as an orbifold projection }
\label{sec:volume-reduction}

Consider a $d$-dimensional $U(\Nc)$ gauge theory,
with or without matter fields,
formulated in a
finite spacetime volume $\Lambda$ taken, for simplicity,
to be a periodic hypercube
of volume $L^d$.%
\footnote
    {
    Generalizing this discussion to the case of
    anisotropic volumes, or mappings which shrink different
    dimensions by different amounts, is completely straightforward.
    For ease of notation, we will stick to simple cubic volumes
    in our discussion.
    }
Because of the periodic boundary conditions,
the translation invariance of the theory is
a compact $U(1)^d$ global symmetry.
Choose any integer $K > 1$, and consider the discrete $(\Z_K)^d$
subgroup of the translation symmetry generated by translations
through multiples of the distance $L' \equiv L/K$
along any coordinate direction.
Let ${\cal T}_K$ denote this discrete translation group,
and let $\Lambda'$ denote a periodic hypercube of volume $(L')^d$.

The reduction of the theory from the volume $\Lambda$ to the smaller
volume $\Lambda'$ may naturally be viewed as an orbifold
projection in which one eliminates from the theory all
degrees of freedom which are not invariant under the
translation subgroup ${\cal T}_K$.
For the criterion of `non-invariance' to be meaningful,
one must first choose variables which transform irreducibly under
the symmetry subgroup defining the projection.
In this case, that just means Fourier transforming fields from
spacetime to momentum space.
In the original volume $\Lambda$, momenta are quantized in multiples
of $2\pi/L$,
so for some generic field $\Phi$,
\begin{equation}
    \Phi(x) = \sum_{n \in \Z^d} \Phi_n \> e^{2\pi i n \cdot x/L} \,,
\label{eq:Fourier-series1}
\end{equation}
where $n$ is an integer-valued $d$-component vector.%
\footnote
    {
    For gauge fields, we assume
    a choice of gauge which preserves periodic boundary conditions.
    Or one may consider a lattice gauge theory without gauge fixing,
    in which case $L$ (divided by the lattice spacing, assumed to be
    set to one) must be multiple of $K$,
    and (\ref{eq:Fourier-series1})
    will have a cutoff of $L$ on the components of $n$,
    for both link and site variables.
    If any fields (such as fermions) are defined with antiperiodic
    boundary conditions, then their allowed momenta in volume $\Lambda$
    will have components of the form $(2n{+}1)\pi/L$.
    In this case, $K$ must be odd so that
    the allowed momenta in the smaller volume $\Lambda'$ are
    commensurate with allowed momenta in the larger volume.
    Although
    Eqs.~(\ref{eq:Fourier-series1})--(\ref{eq:trans-invariance})
    show the case of periodic fields, their extension to
    antiperiodic fields is obvious.
    }
Eliminating degrees of freedom which are not invariant under
the translation subgroup ${\cal T}_K$ simply means setting to zero
all Fourier coefficients $\Phi_n$ for which the components of $n$
are not integer multiples of $K$.
This is the same as averaging $\Phi(x)$ over all translations along
coordinate axes by multiples of $L'$.
Consequently, the projection takes
\begin{equation}
    \Phi(x) \longrightarrow
    \widetilde
    \Phi(x) \equiv \sum_{m \in \Z^d}
    \widetilde\Phi_m \> e^{2\pi i m \cdot x/L'} \,,
\label{eq:Fourier-series2}
\end{equation}
where $\widetilde\Phi_m = \Phi_{mK}$.
This is the Fourier series for a field in the smaller
periodic volume $\Lambda'$.
The net effect of this projection is to replace every field,
which was initially periodic with period $L$, by a projected field
which is periodic with period $L'$.
In other words, the projection is the same as
imposing, on every field of the theory,
a constraint of shorter periodicity,
\begin{equation}
    \Phi(x+L' \, \hat e_\nu) = \Phi(x) \,,
\label{eq:trans-invariance}
\end{equation}
where $\{ \hat e_\nu \}$ denote coordinate basis vectors.
We have belabored the discussion of this simple projection
in order to emphasize that ``projecting out'' degrees of freedom
which are not invariant under some symmetry subgroup is exactly
equivalent to imposing a set of constraints,
such as Eq.~(\ref{eq:trans-invariance}), expressing invariance
of fields under the chosen symmetry transformations.
This is equally true for all other orbifold projections.

The projection (\ref{eq:Fourier-series2}),
applied to all fields of the theory,
may be regarded as defining a mapping between
observables of the ``parent'' theory, defined in the volume $\Lambda$,
and the ``daughter'' theory, defined in $\Lambda'$.
Under this mapping, all observables which carry momenta that are
not quantized in units of $2\pi/L'$ map to zero.

In theories without fundamental representation matter fields
(but possibly with adjoint or rank-two tensor representation fields)
the natural gauge invariant observables are ``single-trace'' operators
which are Wilson loops, or Wilson loops decorated with arbitrary insertions
of matter fields along the loop.
Such single-trace observables may be classified by their net winding
numbers around each cycle of the periodic volume in which the theory
is defined.%
\footnote
    {
    In $U(N)$ theories with only adjoint matter fields,
    winding numbers are signed integers.
    In theories with rank-two symmetric or antisymmetric tensor
    representation matter,
    one should regard winding numbers as unsigned integers
    because gauge invariant single-trace
    observables such as
    $
	\tr (U[C_{x\to y}] \Phi_y U[C'_{y\to x}]^* \Phi_x^*)
    $
    cannot be assigned a consistent orientation.
    ($U[C_{x\to y}]$ denotes a parallel transporter along a
    curve $C$ running from point $x$ to $y$.)
    }
The mapping (\ref{eq:Fourier-series2})
defines a one-to-one correspondence between
gauge invariant single-trace observables in the parent theory
which are invariant under ${\cal T}_K$
({\em i.e.}, observables averaged over all translations by multiples of $L'$)
and single-trace observables in the daughter theory
whose winding numbers are integer multiples of $K$.
If fundamental representation fields are present,
then the mapping gives a one-to-one correspondence between
mesonic observables in the parent which are invariant under $\mathcal T_K$
and arbitrary mesonic observables in the daughter theory.
These classes of gauge-invariant observables, in both parent and daughter,
will be termed ``neutral''.

Since the action of the parent theory contains a spacetime integral
over the volume $\Lambda$,
which is $K^d$ times the volume of $\Lambda'$,
the projection (\ref{eq:Fourier-series2}) does not map
the action of the parent theory to
the action of the daughter theory.
Rather,
\begin{equation}
    S_{\rm parent} \longrightarrow K^d \> S_{\rm daughter} \,.
\label{eq:Smap1}
\end{equation}
The prefactor is the ratio of the number of degrees
of freedom in parent and daughter theories.
This form of the relation between parent and daughter actions,
involving a rescaling by an overall multiplicative factor,
is common to all orbifold projections
but has, on occasion, been misunderstood
(as noted in Ref.~\cite{KUY3}).

For a lattice gauge theory
(on a simple cubic lattice, with the lattice spacing set to one),
if one chooses $K = L$ then
this volume-reducing projection
maps the parent theory defined on an $L^d$ site lattice
to a daughter theory defined on a lattice with just one site.
For the specific case of pure Yang-Mills theory
with the usual Wilson action,
\begin{equation}
    S_{\rm W}
    \equiv
    \frac{\Nc}\lambda \sum_{x\in (\Z_L)^d} \> \sum_{\mu < \nu} \>
    \tr \>\big(
	U_\mu[x] \,
	U_\nu[x{+}\hat e_\mu] \,
	U_\mu^\dagger[x{+}\hat e_\nu] \,
	U_\nu^\dagger[x]
	+ \mbox{h.c.}
	\big) \,,
\label{eq:Wilson-action}
\end{equation}
the result is the Eguchi-Kawai model,
which is a $U(\Nc)$ matrix model with $d$ independent unitary matrices
$\{ U_\nu \}$ and action
\begin{equation}
    S_{\rm EK}
    \equiv
    \frac\Nc\lambda \; \sum_{\mu < \nu} \>
    \tr \>
    \big(U_\mu \,U_\nu \,U_\mu^\dagger \,U_\nu^\dagger + \mbox{h.c.}\big)\,.
\label{eq:EKaction}
\end{equation}
Here $\lambda \equiv g^2 \Nc$ is the (bare) 't Hooft coupling,
which is held fixed as $\Nc \to \infty$.
(The conventional lattice coupling $\beta \equiv 2 \Nc^2/\lambda$.)

\subsection{Volume expansion as an orbifold projection}

To represent a volume-increasing mapping as an orbifold projection,
it is easiest to use the language of lattice gauge theory.%
\footnote
    {
    As discussed below, volume enlarging projections involve
    the center symmetry of the theory
    in an essential fashion.
    In a lattice formulation of the theory (on a cubic lattice)
    this invariance conveniently appears as a simple global symmetry
    corresponding to phase rotations of all link variables pointing
    in a given direction.
    }
Before considering more general examples,
it is useful to start with the simplest case of
the single-site $U(\Nc)$ Eguchi-Kawai model.
(Volume-enlarging projections starting from an arbitrary size lattice
are discussed in the Appendix.)
The addition of matter fields will be considered below.

Choose an integer $K > 1$ and let $\Nc$ be divisible by $K^d$,
so that $\Nc = K^d N$ for some integer $N$.
We wish to show that a $(\Z_K)^d$ orbifold projection can map this theory
into a $d$-dimensional $U(N)$ pure gauge theory on a periodic $K^d$
size lattice.%
\footnote
    {
    Generalizing the following treatment to the case
    of a $\Z_{K_1} \times \Z_{K_2} \times \cdots \times \Z_{K_d}$
    projection mapping the EK model to a pure gauge theory on
    a periodic lattice of size $K_1 \times K_2 \times \cdots \times K_d$
    is straightforward.
    }

The Eguchi-Kawai (EK) model is invariant under $U(\Nc)$ transformations
which act by conjugation,
\begin{equation}
    U_\mu \to V \, U_\mu \, V^\dagger \,,\qquad V \in U(\Nc) \,.
\label{eq:EKsym1}
\end{equation}
This is the reduction to one site of the usual action of
gauge transformations on link variables,
$
    U_\mu[x] \to V[x] \, U_\mu[x] \, V[x{+}\hat e_\mu]^\dagger
\label{eq:EKsym2}
$.
The EK model is also invariant under a $U(1)^d$ symmetry which
multiplies each matrix by an arbitrary phase,
\begin{equation}
    U_\mu \to e^{i\alpha_\mu} \, U_\mu \,.
\label{eq:EKsym3}
\end{equation}
This is the reduction to one site of what is commonly called
center symmetry, which is the invariance of gauge theories
with periodic boundary conditions
(and only adjoint representation matter fields)
under gauge transformations which are not themselves periodic,
but are periodic up to some element of the center of the gauge group
\cite{center-sym1,center-sym2}.

The required projection will select degrees of freedom that are invariant
under a $(\Z_K)^d$ symmetry subgroup which is embedded non-trivially
within the full $U(1)^d \times U(\Nc)$ symmetry group.
The net effect of the projection is to impose a set of constraints
having the form
\begin{equation}
    U_\mu
    =
    \begin{cases}
    \gamma_\nu \, U_\mu \, \gamma_\nu^\dagger \;
    e^{2\pi i/K} \,,
    & \mu = \nu \,;
    \\
    \, \gamma_\nu \, U_\mu \, \gamma_\nu^\dagger \,,
    & \mu \ne \nu \,,
    \end{cases}
\label{eq:EKconstraint}
\end{equation}
for all $\mu,\nu$,
with $\{ \gamma_\nu \}$ a particular set of $d$ mutually commuting
$(K^dN) \times (K^dN)$ unitary matrices whose eigenvalues are all $K$-th
roots of unity, each with multiplicity $K^{d-1}N$.
An explicit definition of these matrices is given in the Appendix.

If one block-decomposes the matrix $U_\mu$ into $K^d \times K^d$ blocks,
each of which is $N \times N$, then the constraint (\ref{eq:EKconstraint})
eliminates all but $K^d$ of these blocks.
The unitarity condition for the full matrix, $U_\mu U_\mu^\dagger = 1$,
implies that each one of the surviving blocks must be an
$N \times N$ unitary matrix.
The constraint reduces the $U(\Nc)$ symmetry (\ref {eq:EKsym1})
of the EK model to the smaller product group $[U(N)]^{K^d}$ ---
which is the full gauge group for a $U(N)$ lattice gauge
theory on a $K^d$ site lattice.
Each surviving block of $U_\mu$ may be naturally associated%
\footnote
    {
    The appropriate association of surviving blocks in $U_\mu$ with links
    in the lattice $\Lambda$ is spelled out in detail
    (and in more generality) in the Appendix.
    But the basic idea is simple:
    Arbitrarily pick a surviving block of $U_1$
    and some site $x$ of the lattice $\Lambda$,
    and map the chosen block of $U_1$ to the link extending from
    site $x$ in the $x$-direction.
    The term $\tr (U_1 U_2 U^\dagger_1 U^\dagger_2)$
    in the EK model action (\ref{eq:EKaction})
    couples the chosen block of $U_1$ to precisely one surviving block of $U_2$
    (and one block of $U_1^\dagger$ and one block of $U_2^\dagger$).
    Map these blocks to the links forming the plaquette in the $x$-$y$
    plane which starts at site $x$.
    Continuing in the same fashion, considering plaquettes
    containing one or more already-assigned links, leads to a
    one-to-one mapping from surviving blocks of the matrices $\{ U_\mu \}$ to
    links of the lattice $\Lambda$.
    }
with an individual link pointing in the $\mu$ direction of a periodic
cubic lattice $\Lambda$ of size $K^d$, in such a fashion
that the EK model action (\ref{eq:EKaction}),
evaluated with matrices satisfying the constraints (\ref{eq:EKconstraint}),
becomes precisely the Wilson action (\ref{eq:Wilson-action})
for the lattice $\Lambda$ and gauge group $U(N)$.
The product of $U(\Nc)$ Haar measures of the EK model, $\prod_\mu dU_\mu$,
similarly reduces to the the product of Haar measures,
$\prod_{x,\mu} dU_\mu[x]$,
appropriate for a $U(N)$ gauge theory on $\Lambda$.

Under this mapping, every single-trace observable of the EK model
whose winding numbers are integer multiples of $K$
({\em i.e.,} where
the net number of $U_\mu$ minus $U_\mu^\dagger$ matrices,
for each $\mu$, are zero mod $K$)
is mapped to a Wilson loop on $\Lambda$ averaged over all translations,
\begin{equation}
    \frac 1\Nc \>
    \tr \left(U_{\mu_1} \, U_{\mu_2} \, U_{\mu_3} \cdots \right)
    \longrightarrow
    \frac 1{|\Lambda|}
    \sum_{x \in \Lambda} \>
    \frac 1N \,
    \tr \left(
	U_{\mu_1}[x] \,
	U_{\mu_2}[x{+}\hat \mu_1] \,
	U_{\mu_3}[x{+}\hat \mu_1{+}\hat\mu_2] \cdots
    \right) \,,
\end{equation}
where $|\Lambda| = K^d$ is the number of sites of the lattice $\Lambda$.
And every single-trace observable with non-zero winding (mod $K$) maps to zero
under the projection (\ref{eq:EKconstraint}).

Because of the overall factor of the rank of the gauge group
in the actions (\ref{eq:Wilson-action}) and (\ref{eq:EKaction}),
this projection from the $U(K^d N)$ EK model to a $U(N)$ gauge theory
on lattice $\Lambda$
maps the EK action to $K^d$ times
the Wilson action on $\Lambda$,
so once again
\begin{equation}
    S_{\rm parent} \longrightarrow K^d \, S_{\rm daughter} \,,
\label{eq:Smap2}
\end{equation}
(with the actions on both sides defined with coinciding values
of the 't Hooft coupling).

As discussed in the Appendix,
this volume-enlarging orbifold projection may be generalized
to the case of an initial periodic lattice $\Lambda'$ of arbitrary
size $(L)^d$ and a final periodic lattice $\Lambda$ of size
$(KL)^d$, for any integer $K > 1$.
Therefore, it is completely natural to view these volume-enlarging
orbifold projections as the ``inverses'' of the volume-reducing projections
discussed in the previous subsection.

\subsubsection{Adjoint representation matter fields}

Generalizing these volume-enlarging projections to
theories with adjoint representation matter fields
(which may be either fermions or scalars)
is straightforward.
Consider, for example, adding an additional Hermitian matrix $\Phi$
to the one-site EK model,
with action
\begin{equation}
    S_{\rm parent} = S_{\rm EK}
    + \Nc \> \tr \Bigl[
    \kappa \sum_\mu \Phi \, U_\mu \, \Phi \, U_\mu^\dagger + V(\Phi)
    \Bigr]\,.
\label{eq:SEK+matter}
\end{equation}
The hopping parameter $\kappa$ and the scalar potential $V(\Phi)$
(which is some univariate function, bounded from below)
are to be held fixed as $\Nc\to\infty$.
Note that the $U(1)^d$ transformations (\ref {eq:EKsym3})
remain a symmetry in the presence of adjoint representation matter fields.
The specific $(\Z_K)^d$ symmetry transformations which define the projection
of interest act on $\Phi$ by conjugation, so the constraints
(\ref{eq:EKconstraint}) on $\{ U_\mu \}$ are simply augmented
by the conditions
\begin{equation}
    \Phi = \gamma_\nu \, \Phi \, \gamma_\nu^\dagger \,,
    \qquad \nu = 1,{\cdots}, d \,,
\label{eq:Phi-constraint}
\end{equation}
for the matrix $\Phi$.
The net effect of the constraint (\ref{eq:Phi-constraint})
is to reduce $\Phi$ to block diagonal form,
with $K^d$ blocks, each of which is $N \times N$.
Each one of these blocks may be associated with a site of the
lattice $\Lambda$ in such a way that the result is a conventional
$U(N)$ lattice gauge theory with an adjoint representation
Hermitian scalar field.
More precisely, the action (\ref{eq:SEK+matter}) of the parent theory
is mapped to $K^d$ times the daughter theory action
[{\em cf.} (\ref {eq:Smap2})], with
\begin{eqnarray}
    S_{\rm daughter}
    &=&
    S_{\rm W}
    + N \sum_{x\in \Lambda}
    \tr\,\Bigl\{
    \kappa \sum_\mu
    \Phi[x] \, U_\mu[x] \, \Phi[x{+}\hat\mu] \, U_\mu^\dagger[x]
    + V(\Phi[x])
    \Bigr\} \,.
\end{eqnarray}
As in the pure gauge case, there is a one-to-one mapping between
single trace observables in the parent matrix model whose winding numbers
are zero modulo $K$, and single trace observables in the daughter lattice
theory which are averaged over lattice translations.%
\footnote
    {
    If the action of the starting theory contains
    multi-trace operators, then the direct application of
    a volume-enlarging projection leads to a non-local daughter theory
    whose action contains terms involving multiple integrals
    (or lattice sums) over spacetime.
    However, such non-local terms may be eliminated by adding
    to the action terms of the form
    $
	\sum_{x,y} [ \O(x) -\O(y) ]^2
    $
    involving the zero momentum variance (or higher cumulants)
    of gauge invariant operators.
    Adding such terms
    does not affect the leading large $\Nc$ dynamics,
    within the neutral sector,
    provided translation invariance is not spontaneously broken.
    }

\subsubsection{Fundamental representation matter fields}

Defining an analogous volume-enlarging projection for
theories with fundamental representation matter fields is,
at first sight, problematic since fundamental representation fields
break the $U(1)^d$ center symmetry which was an essential ingredient
in the projection (\ref {eq:EKconstraint}).
However, this is not an insurmountable obstacle for theories with
a $U(\Nf)$ flavor symmetry.
One may choose to regard a theory with a global flavor symmetry
as the zero-coupling limit of a theory in which the $U(\Nf)$
flavor symmetry is weakly gauged.
In other words, a theory with a $U(\Nc)$ gauge group
and fundamental representation matter with $U(\Nf)$ flavor symmetry
may equally well be regarded as a $U(\Nc) \times U(\Nf)$ gauge theory with
bifundamental matter, in the limit of vanishing coupling for the
$U(\Nf)$ gauge group.
The resulting $U(\Nc) \times U(\Nf)$ theory is precisely a quiver
gauge theory of the class discussed in section \ref{sec:quiver}.
This reinterpretation is helpful because the
$U(\Nc) \times U(\Nf)$ theory, with bifundamental matter,
now contains a $U(1)^d$ center symmetry
(which transforms the $U(\Nc)$ and $U(\Nf)$ gauge fields
just like the pure Yang-Mills case, and leaves the matter fields invariant).
If $\Nf$ is divisible by $K^d$, so that $\Nf = K^d \, \nf$,
then a volume-enlarging projection may be defined by suitably
embedding the $(\Z_K)^d$ symmetry determining the projection
into the product of the gauge and center symmetry groups of the
$U(\Nc) \times U(\Nf)$ quiver theory.
The resulting constraints have the effect of reducing
the matter fields, viewed as an $\Nc \times \Nf$ matrix,
to a block-diagonal form with $K^d$ blocks each of which is $N \times \nf$.
If $\Nf/\Nc$ is held fixed as $\Nc \to \infty$, then this
volume-enlarging projection may again naturally be viewed as
the ``inverse'' of the volume reducing projection.

\subsubsection{Tensor representation matter fields}

Adding rank-two symmetric or antisymmetric tensor representation fields
to a $U(\Nc)$ Yang-Mills theory reduces the center symmetry from
$U(1)^d$ to $(\Z_2)^d$ (corresponding to negating all link variables pointing
in a given direction.)
Starting from a single-site model, one may use this symmetry
(plus gauge invariance) to define a volume enlarging projection
which maps the theory to one on a periodic cubic lattice of size $2^d$.
More generally, if one starts with the theory on a periodic lattice
of size $L^d$, one may define volume enlarging projections which
double the length of one or more lattice directions --- provided $L$ is odd.
But, for reasons discussed in the Appendix,
the same approach does not work if $L$ is even.%
\footnote
    {
    In brief, if $L$ is even then the result of the projection is
    not a theory on a larger lattice,
    but rather multiple decoupled copies of the
    theory on the original lattice.
    See the Appendix for details.
    }

However, just as with fundamental representation fields,
if one considers a theory with $\Nf$ degenerate flavors of matter fields,
then a $U(\Nc)$ gauge theory with a global $U(\Nf)$ flavor symmetry
may be viewed as the zero (flavor) coupling limit of a
$U(\Nc) \times U(\Nf)$ gauge theory --- which does have a $U(1)^d$
center symmetry.
And with this enlarged gauge group one may define projections which
increase the volume by $K^d$, with $K$ any integer, provided
both $\Nc$ and $\Nf$ are divisible by $K^d$.%
\footnote
    {
    However,
    with tensor representation matter fields
    it should be noted that
    one cannot define a natural large $\Nc$ limit by
    sending $\Nc \to \infty$ with $\Nf/\Nc$ held fixed.
    Consequently, this way of defining a volume enlarging projection
    will not be useful to us.
    }

\subsection {Large $\Nc$ equivalence}
\label{sec:largeN}

In Refs.~\cite{KUY1,KUY2},
we proved an equivalence between the large $N$ limits of
a wide class of $U(N)$ field theories with adjoint matter
and their orbifold projections yielding quiver gauge theories.
More precisely, we demonstrated that the large-$N$ dynamics
of the parent and daughter theories,
within their respective neutral sectors, coincide.%
\footnote
    {
    In the parent theory, neutral states (or operators)
    are those which are invariant under the symmetries used
    to define the projection.
    In the daughter theory, neutral states (or operators)
    are those which are invariant under global symmetries in the daughter
    which are remnants of gauge symmetries in the parent theory.
    Non-neutral operators are also called ``twisted''.
    }
If the symmetries defining the neutral sectors are not spontaneously
broken, then the ground states of parent and daughter theories will lie
within their respective neutral sectors.
In this case, the expectation values of corresponding single-trace neutral
operators in parent and daughter theories will have coinciding large $N$ limits.
Moreover, the leading large-$N$ behavior of {\em connected} correlators
(as well as ground state or thermal free energies)
are also directly related.%
\footnote
    {
    If the projection maps the neutral single trace operators $\O_i$
    in the parent to corresponding operators $\Otilde_i$ in the daughter,
    then
    \begin{equation}
	\lim_{|G_p|\to\infty}
	|G_p|^{M-1}
	\langle \O_1 \cdots \O_M \rangle_{\rm conn}
	=
	\lim_{|G_d|\to\infty}
	|G_d|^{M-1}
	\langle \Otilde_1 \cdots \Otilde_M \rangle_{\rm conn} \,,
    \label{eq:connected-corr}
    \end{equation}
    where $|G_p|$  and $|G_d|$ denote the dimension of the gauge group
    in the parent and daughter theory, respectively.
    Ground state energies (or free energies)
    satisfy the relation,
    \begin{equation}
	\lim_{|G_p|\to\infty}
	|G_p|^{-1}
	E_{\rm g.s.}^{\rm parent}
	=
	\lim_{|G_d|\to\infty}
	|G_d|^{-1}
	E_{\rm g.s.}^{\rm daughter} \,.
    \label{eq:Egs}
    \end{equation}
    \label{fn:largeNmapping}
    }
It is important to bear in mind that this large-$N$ equivalence
applies only to observables within the respective neutral sectors
of the parent and daughter theories.

Ref.~\cite{KUY1} demonstrated this large-$N$ equivalence by
comparing generalized loop equations (or Schwinger-Dyson equations
for correlators of gauge invariant operators) of parent and daughter theories.
Ref.~\cite{KUY2} used a more abstract, but more powerful, approach
involving the comparison of large $N$ classical dynamics constructed
from suitable large $N$ coherent states.%
\footnote
    {
    The comparison of loop equations directly reveals the
    necessity of the symmetry realization conditions for large-$N$
    equivalence of the lattice regularized field theories, or
    their continuum limits.
    If the symmetries defining the neutral sectors
    are spontaneously broken,
    then the loop equations of the two theories do not coincide,
    invalidating equivalence.
    Conversely, unbroken symmetries
    defining the neutral sectors imply coinciding loop equations.
    However,
    because loop equations generally can (and do) have multiple solutions,
    a demonstration of coinciding loop equations
    does not constitute a proof of equivalence,
    except in the large coupling, large mass phase of the lattice regulated
    theory where one can prove that the loop equations have
    a unique physical solution.
    The coherent state approach of Ref.~\cite{KUY2} plugs this loophole
    and demonstrates equivalence in any phase of the theories
    (including their continuum limits) which
    satisfy the necessary and sufficient symmetry realization conditions.
    For early literature discussing loop equations in presence of adjoint 
    matter, see also Ref.~\cite{Heller:1982gg}.    
}

The methods, and results, of Refs.~\cite{KUY1,KUY2} extend immediately
to the volume-reducing and volume-enlarging projections
discussed above
in gauge theories containing only adjoint representation matter fields.
(The only difference is the precise form of the projection, and every step of
the analysis of Refs.~\cite{KUY1,KUY2} goes through with only minor
cosmetic changes.
Therefore, we refer readers to these references for details.)
Consequently, the volume-reducing and volume-enlarging projections
produce theories with coinciding large-$N$ limits within their
neutral sectors, just as discussed above.%
\footnote
    {
    For a $U(\Nc)$ gauge theory on a lattice of size $L^d$,
    note that the actual gauge symmetry group is $U(\Nc)^{L^d}$
    (since there is an independent $U(\Nc)$ group on each site)
    with a total dimension of $L^d \, \Nc^2$.
    For a volume-reducing projection mapping a parent $U(\Nc)$ theory
    on a lattice $\Lambda$ of size $(KL)^d$ to a daughter $U(\Nc)$ theory
    on a lattice $\Lambda'$ of size $L^d$,
    the factor in the relation (\ref{eq:Smap1}) between
    the actions of these theories equals the ratio of the dimensions
    of their gauge groups,
    $|G_p|/|G_d| = K^d$.
    And for the volume-enlarging projection taking
    a $U(K^d N)$ theory on a size $L^d$ lattice to a $U(N)$ theory
    on a size $(KL)^d$ lattice,
    the factor in the relation (\ref{eq:Smap2}) between
    the actions in this case is also the ratio of the dimensions
    of their gauge groups,
    $|G_p|/|G_d| = [L^d (K^d N)^2]/[(KL)^d N^2] = K^d$.
    Therefore, these mappings between actions could have been
    written in either case as
    $
	|G_p|^{-1} S_{\rm parent} \to |G_d|^{-1} S_{\rm daughter}
    $,
    which is precisely the relation (\ref{eq:Egs})
    between actions (or ground state energies)
    noted in footnote \ref{fn:largeNmapping}.
    }
The corresponding situation for theories with fundamental
or tensor representation fields is discussed separately below.

If $\{\O_i\}$ are arbitrary gauge invariant single trace operators,
with definite momenta which are commensurate with an $L^d$ volume
({\em i.e.}, whose momentum components are integer multiples of $2\pi/L$),
then the large-$N$ equivalence applied to the
the projection from a volume $\Lambda = (KL)^d$
to a smaller volume $\Lambda' = (L)^d$
implies that connected correlators of these operators satisfy
\begin{equation}
    \lim_{N\to\infty}
    (K^d N^2)^{M-1}
    \langle \O_1 \cdots \O_M \rangle_{\rm conn}^{N,KL}
    =
    \lim_{N\to\infty}
    (N^2)^{M-1}
    \langle \O_1 \cdots \O_M \rangle_{\rm conn}^{N,L} \,,
\label{eq:connected-corr2}
\end{equation}
where $\langle \cdots \rangle^{N,L}$ denotes an expectation value
in a periodic cube of size $L$ with the specified value of $N$,
{\em provided} the theory in volume
$\Lambda'$ does not spontaneously break the $(\Z_K)^d$ subgroup
of the $U(1)^d$
center symmetry and the theory in volume $\Lambda$
does not spontaneously break translation invariance (by multiplies of $L$).
Similarly, the large-$N$ equivalence applied to the volume-enlarging
projection from $\Lambda'$ back to $\Lambda$ implies that
\begin{equation}
    \lim_{N\to\infty}
    (N^2)^{M-1}
    \langle \O_1 \cdots \O_M \rangle_{\rm conn}^{N,KL}
    =
    \lim_{N\to\infty}
    (K^d N^2)^{M-1}
    \langle \O_1 \cdots \O_M \rangle_{\rm conn}^{K^d N,L} ,
\label{eq:connected-corr3}
\end{equation}
again provided that $(\Z_K)^d$ center symmetry is not spontaneously broken
in volume $\Lambda'$,
and translation invariance is unbroken in volume $\Lambda$.

Volume-changing orbifold projections directly connect
theories in volumes of commensurate size
({\em i.e.}, with spatial periods which are integer multiples of each other).
Volume independence is more general,
but this too follows from the large $N$ orbifold equivalence,
because the equivalence of large $N$ dynamics
(within respective neutral sectors)
is a transitive relation among theories.
For example,
consider a $U(\Nc)$ gauge theory
formulated in volume $\Lambda = (KL)^d$
and in volume $\Lambda' = (K'L)^d$.
Let $Q$ be the greatest common divisor of $K$ and $K'$,
$Q = \gcd(K,K')$.
If $K$ is not divisible by $K'$ (or vice-versa), then no
volume-changing projection directly relates these two volumes.
However, a volume-enlarging projection relates both of these
theories to volume $\Lambda'' = (K''L)^d$, with $K'' = K K' / Q$.
Therefore, correlators of corresponding neutral observables in the
theories in volume $\Lambda$ and $\Lambda'$ satisfy
the appropriate generalization of Eq.~(\ref{eq:connected-corr3}),
namely,
\begin{equation}
    \lim_{N\to\infty}
    (K'^d N^2)^{M-1}
    \langle \O_1 \cdots \O_M \rangle_{\rm conn}^{(K'/Q)^d N,KL}
    =
    \lim_{N\to\infty}
    (K^d N^2)^{M-1}
    \langle \O_1 \cdots \O_M \rangle_{\rm conn}^{(K/Q)^d N,K'L} ,
\label{eq:connected-corr4}
\end{equation}
provided the $(\Z_{K'/Q})^d$ center symmetry is not spontaneously
broken in volume $\Lambda$, and the
$(\Z_{K/Q})^d$ center symmetry is not broken in $\Lambda'$.
This relation is applicable to single-trace observables
whose winding numbers vanish mod $K'/Q$ in $\Lambda$
(or mod $K/Q$ in $\Lambda'$),
and whose momenta are quantized in units of $2\pi/(QL)$.

Of course, the recognition of the volume independence of large $N$
gauge theories is far from new and, in the case of pure gauge theories,
goes back to the original paper of Eguchi and Kawai \cite {Eguchi-Kawai}.
(The coherent state proof of equivalence, for pure gauge theories,
is contained in a small comment in Ref.~\cite{LGY-largeN}.)
But the natural connection to orbifold projections,
the appropriate extension to theories with matter,
and the validity of the above relations
for connected correlators%
\footnote{
   The large-$N$ equivalence of the Eguchi-Kawai matrix model
   (\ref{eq:EKaction}) to lattice gauge theory only applies to
   neutral operators $\O_i$ in the finite-volume theory with
   zero momentum.
   Therefore, information (such as glueball masses)
   encoded in long-distance behavior of correlations is not captured by
   the matrix model.
   Retaining such information in a reduced volume theory requires
   keeping at least one uncompactified dimension.
   See Ref.~\cite{Levine:1982fa} for previous discussion of this
   issue in reduced models.
}
have not been widely appreciated.

\FIGURE[t]{
\begin{minipage}[t]{.48\textwidth}
\begin{center}
  \psfrag{V0}{$V{=}0$}
  \psfrag{Vinf}{$V{=}\infty$}
  \psfrag{b}{$1/\lambda_0$}
  \psfrag{inf}{$\infty$}
  \epsfig{file=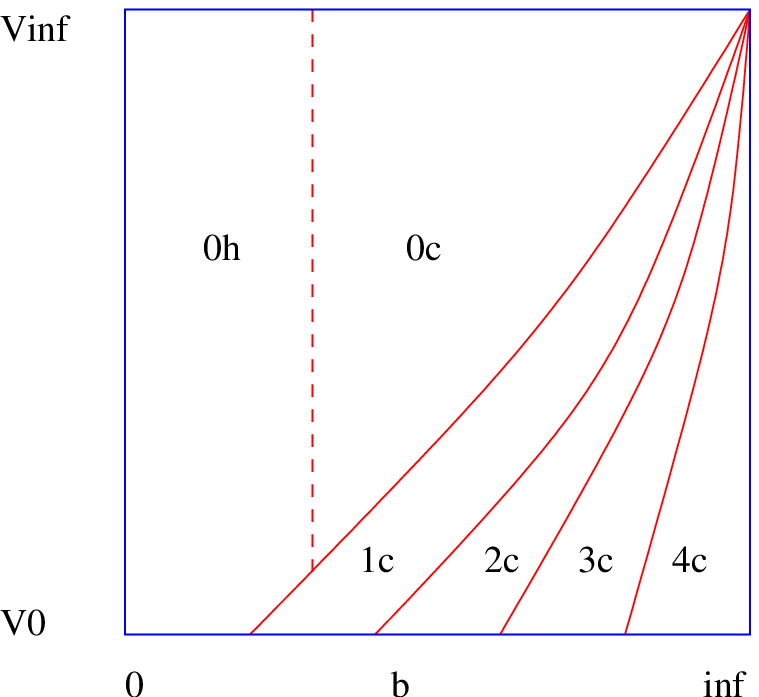,scale=0.9}
\end{center}
\end{minipage}
\begin{minipage}[t]{.48\textwidth}
\begin{center}
  \psfrag{V0}{$V{=}0$}
  \psfrag{Vinf}{$V{=}\infty$}
  \psfrag{b}{$1/\lambda_0$}
  \psfrag{inf}{$\infty$}
  \psfrag{Lc}{$L_c(\lambda_0)$}
  \epsfig{file=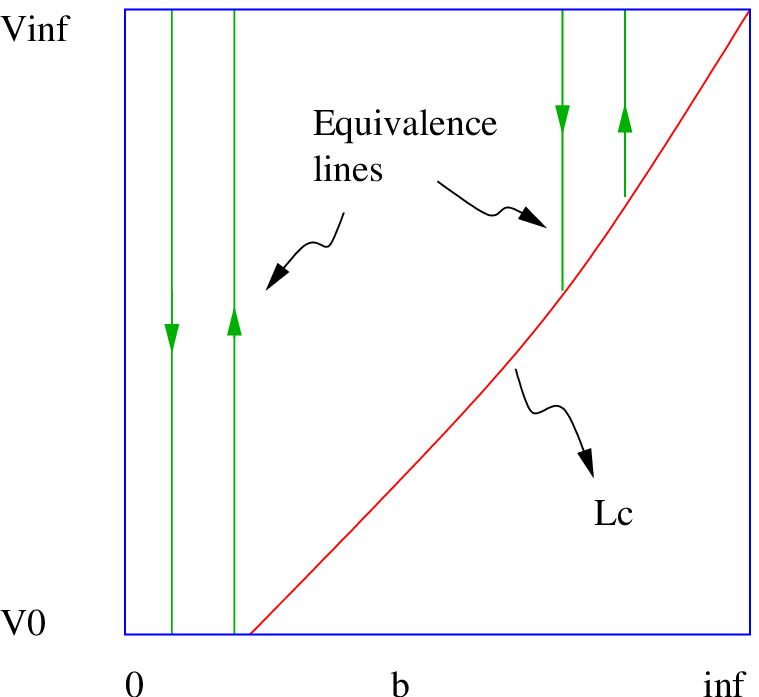,scale=0.9}
\end{center}
\end{minipage}
\caption{
    Left:
    the phase diagram of $U(\Nc)$ lattice Yang-Mills theory
    on a four-dimensional periodic cube,
    in the limit of large $\Nc$,
    as a function the lattice volume $V$
    and inverse (bare) 't Hooft coupling $\lambda_0$.
    The $V=0$ axis corresponds to the single-site Eguchi-Kawai model.
    Solid red lines separate phases with different realizations of the
    $U(1)^4$ center symmetry, as described in the text.
    These lines approach finite physical volumes (and infinite lattice volume)
    in the $\lambda_0\to0$ continuum limit.
    The dashed line is a lattice artifact and represents
    a Gross-Witten type phase transition in which the one plaquette
    eigenvalue distribution develops a gap.
    The figure is adapted from a description in
    Ref.~\cite{Kiskis-Narayanan-Neuberger}.
    Right:
    volume independence of large $N$ lattice Yang-Mills theory,
    as a function of the inverse (bare) 't Hooft coupling.
    The vertical green lines are lines of large $N$ equivalence.
    Both volume reduction (downward arrows) and volume expansion (upward
    arrows) may be interpreted as orbifold projections.
    The red line corresponds to the first center-symmetry breaking
    phase transition (separating the 1c phase from the 0c or 0h phases).
    Volume independence fails below this line.
    As discussed in the text, the addition of light adjoint representation
    matter fields can eliminate the existence of the phases with
    spontaneously broken center symmetry.
} 
\label{fig:phase}
} 

Translation invariance is not expected to break spontaneously
in any normal local field theory.
But, as noted in the Introduction, it has long been known that
for sufficiently weak coupling the single-site EK model
(in more than two dimensions) spontaneously breaks the
$U(1)^d$ center symmetry and therefore its correlators
cease being equivalent (in the large $N$ limit) to those
of large volume Yang-Mills theory.
For large-$N$ pure gauge theory on larger periodic cubic lattices,
the situation has been clarified in recent years thanks to the
numerical work of Kiskis, Narayanan and Neuberger
\cite{Narayanan-Neuberger,Kiskis-Narayanan-Neuberger}.
(See also earlier work \cite{Neuberger:1983xc}.)
The $d\,{=}\,4$ phase diagram they find, as a function the
bare inverse 't Hooft coupling $1/\lambda_0$ and lattice volume $V$,
is shown schematically in Fig.~\ref{fig:phase}.
The continuum limit corresponds to $1/\lambda_0 \to \infty$.
The $U(1)^4$ center symmetry is broken spontaneously to
$U(1)^3$, $U(1)^2$, $U(1)$, and nothing in the phases labeled
1c, 2c, 3c, and 4c, respectively.
All symmetries are unbroken in the phases labeled 0h and 0c.
These two phases are distinguished by the single plaquette
eigenvalue distribution; in phase 0h it has support on the
entire unit circle, while it develops a gap in phase 0c.
The transition between these two phases is an analog of the
Gross-Witten transition in two dimensions \cite{Gross-Witten},
and is not associated with spontaneous breaking of any symmetry.
The symmetry breaking phase transition lines,
appearing as red curved lines in the figure,
approach scaling lines which reach finite physical
volume (and infinite lattice volume) in the continuum limit
\cite{Kiskis-Narayanan-Neuberger}.

Consequently, large-$N$ volume independence of Yang-Mills theory
(in $d > 2$) is valid only in the 0c and 0h phases, {\em i.e.},
above the first symmetry-breaking phase transition line,
which defines the minimal size $L_c(\lambda_0)$
of a periodic lattice for which large $N$ volume independence holds.
This is illustrated on the right in Fig.~\ref{fig:phase}.
The addition of adjoint representation matter fields
can affect the existence of a minimal size for which large $N$
volume independence holds.
In particular, one or more flavors of light adjoint representation
fermions, defined with periodic boundary conditions, can entirely
suppress the center-symmetry breaking phase transitions.
This is discussed below in the section \ref{EKforQCD}.
In such theories, large-$N$ volume independence holds
for periodic cubic volumes of any size .%

Once again, it must be emphasized that large-$N$ orbifold equivalence
applies only to suitable ``neutral'' observables.
In the case of volume independence, this means single-trace
observables whose momenta and winding numbers are commensurate
with both volumes one is comparing.
For the validity of large-$N$ orbifold equivalence,
what is significant is not the existence of non-neutral sectors,
but rather the realization of the symmetries which define
the relevant neutral sectors.
This inevitably depends on the detailed dynamics of the theory,
as well as the volume under consideration.
The example of large-$N$ volume independence,
depicted in Fig.~\ref{fig:phase},
is a particularly clear illustration of this point.

\subsubsection*{Fundamental representation matter}

As noted earlier, for theories with fundamental representation matter
fields and $U(\Nf)$ flavor symmetry, one can define
a volume-enlarging projection
(for suitably large values of $\Nc$ and $\Nf)$
by regarding the flavor symmetry as weakly gauged and
exploiting the $U(1)^d$ center symmetry which is part of the
enlarged gauge group.
If $\Nf/\Nc$ is held fixed as $\Nc \to \infty$, then
the methods of Refs.~\cite{KUY1,KUY2} also apply to the comparison
of these theories in different volumes,
and imply that large $N$ dynamics coincides
within their respective neutral sectors.
There is, however, one crucial difference affecting the utility of this result.
In a $U(\Nc) \times U(\Nf)$ gauge theory,
the center symmetry which defines the neutral sector of the
smaller volume theory is {\em always\/} spontaneously broken
in the limit of arbitrarily weak coupling for the flavor group.
This merely reflects the fact that
arbitrary constant, commuting values of the flavor gauge field
are equally valid vacuum configurations
in the limit of vanishing gauge coupling for the flavor group.
Therefore large $N$ volume independence does not hold for
vacuum (or thermal) expectation values in theories with
fundamental representation matter when $\Nf/\Nc$ is held fixed.

This failure of large $\Nc$ volume independence may be understood
more physically by considering the Casimir energy of these theories.
In a confining theory, if one or more dimensions are compactified
then the theory will have a non-zero Casimir energy whose value
depends on the size of the compactification.
In a pure gauge theory,
the spectral density of glueballs remains $O(1)$ in the large $\Nc$ limit,
and hence their contribution to the Casimir energy is also $O(1)$.
Large $\Nc$ volume independence only applies to $O(\Nc^2)$ contributions
to the ground state energy, and this is unaffected by the volume
dependent Casimir energy contributions.
The same result holds for theories containing adjoint representation
matter fields (whose number does not grow with $\Nc$).
But if the theory contains $\Nf$ degenerate fundamental representation
matter fields,
then the hadron spectrum will include bound mesons,
and their spectral density will be proportional to $\Nf^2$
(since one can independently specify the flavor of the valence quark
and antiquark).
Consequently, the Casimir energy will scale as $\Nf^2$,
implying that large $\Nc$ volume independence (of ground state properties)
must fail if $\Nf/\Nc$ is held fixed.

If instead one chooses to hold $\Nf$ fixed as $\Nc \to \infty$,
then the fundamental representation fields have no effect
whatsoever on the leading $O(\Nc^2)$ contribution to the
ground state energy (or free energy),
or on the leading large $\Nc$ behavior of connected correlators
of single-trace (gluonic) observables.
However, one may consider subleading $O(\Nc)$
contributions to the vacuum energy (or free energy),
as well as the behavior of connected correlators involving mesonic operators.
Large $\Nc$ volume independence {\em does} hold for $O(\Nc)$
thermodynamic contributions, and for mesonic correlators,
when $\Nf$ is held fixed.
One may show this by examining the coherent state algebra
for the mesonic sector \cite{LGY-largeN2,LGY-largeN3} or,
perhaps more directly, by integrating out the fundamental
representation fermions and considering the resulting
non-local gluonic observables.
Under this procedure,
the expectation value of a fermion bilinear $\bar\psi(x) M \psi(x)$
becomes
$ \tr \, [M G(x,x)]$
where $G(x,y)$ is the Green's function for the Dirac operator
$\rlap {\,\slash} D$ in an arbitrary background gauge field and
$M$ is some matrix acting on Dirac indices.
This matrix element of the Dirac propagator may be
expanded as a linear combination of Wilson loops passing through
the point $x$.
These will include both topologically trivial loops which lie
in the neutral sector of a volume-enlarging projection,
and topologically non-trivial loops with non-zero net winding
around the periodic volume.
However, as long as the center symmetry
of the gluonic theory (in the absence of fundamental representation matter)
is unbroken,
then the expectation value of all topologically non-trivial loops,
which lie in the twisted sector, will vanish.
So the expectation value of any gauge invariant fermion bilinear is
equivalent to the expectation value of a neutral single-trace gluonic operator,
provided the symmetry realization necessary for large $\Nc$ volume
independence is satisfied.

The same argument can be applied to connected correlators of
mesonic operators.
Integrating out the fermions in this case will generate both
expectation values of
single trace contributions of the form
\begin{equation}
    \tr [ M_1 \, G(x_1,x_2) \, M_2 \, G(x_2,x_3) \, M_3 \cdots G(x_M, x_1) ] \,,
\end{equation}
and connected correlators of multi-trace contributions, such as
\begin{equation}
    \tr [ M_1 \, G(x_1,x_1)] \>
    \tr [ M_2 \, G(x_2,x_3) \, M_3 \cdots G(x_M, x_2) ] \,.
\end{equation}
The single-trace contributions are linear combinations of
Wilson loops.
Once again, only the topologically trivial (neutral) loops,
to which large $\Nc$ volume independence applies,
have non-vanishing expectation values
as long as the center symmetry of the gluonic theory is unbroken.
But the multi-trace contributions lead to connected correlators of
both topologically trivial and non-trivial Wilson loops.
Connected correlators of topologically non-trivial Wilson loops
are non-zero even when center symmetry is unbroken.

Despite the existence of these non-neutral (and volume dependent)
contributions to connected correlators of mesonic operators,
volume independence is valid for the leading large $\Nc$
behavior of connected correlators
of arbitrary mesonic operators,%
\footnote
    {
    Or products of mesonic and neutral single-trace operators.
    }
when the center symmetry of the gluonic theory is unbroken
and $\Nf$ is held fixed as $\Nc \to\infty$.
The key point is that the multi-trace contributions to connected
correlators are suppressed by a power of $\Nf/\Nc$
relative to the single trace contributions.
If $\{ \O_i \}$ are mesonic operators scaled to have finite
large $\Nc$ expectation values,
then standard large $\Nc$ counting \cite{Coleman}
shows that the single-trace contributions to
$\langle \O_1 \cdots \O_M \rangle_{\rm conn}$
are $O(\Nc^{-(M-1)})$.
But because connected correlators of gluonic operators
scale with powers $1/\Nc^2$, not $1/\Nc$,
multi-trace contributions to connected correlators of mesonic
operators are subdominant relative to the leading single-trace
contributions, provided $\Nf/\Nc \ll 1$.

\subsubsection*{Tensor representation matter}

An analysis of either loop equations,
or large $N$ coherent state dynamics,
may be used to examine volume independence
in large $\Nc$ theories with rank-two tensor representation matter fields.
Appropriately adapting the methods of Refs.~\cite{KUY1,KUY2},
one finds that large $\Nc$ volume independence is valid,
just as in theories with adjoint representation matter fields.
In particular,
Eqs.~(\ref{eq:connected-corr2})--(\ref{eq:connected-corr4})
expressing the equivalence between connected correlators
(or expectation values) of neutral single trace observables
in volumes $\Lambda = (KL)^d$ and $\Lambda' = (L)^d$
hold provided that translation invariance is not spontaneously
broken in the larger volume theory, and the $(\Z_2)^d$ center symmetry
is not spontaneously broken in the smaller volume theory.%
\footnote
    {
    There are some interesting subtleties with tensor representation
    matter, related to the fact that the center symmetry is only
    $(\Z_2)^d$, not $U(1)^d$.
    In the loop equations for expectation values of winding number
    zero observables,
    the splitting terms arising from self-intersections
    can generate (after using large $N$ factorization)
    the product of expectations of two smaller loops,
    each of which may have a non-zero winding number.
    Since the possible set of such non-zero winding number loops
    is volume dependent,
    if such (non-neutral) loops have non-zero expectation values,
    this violates large $\Nc$ volume independence.
    In $U(\Nc)$ gauge theories with only adjoint representation matter,
    the $U(1)^d$ center symmetry, if unbroken,
    guarantees (for any value of $\Nc$)
    that all loops with non-zero winding numbers have vanishing
    expectation values.
    But in theories with only a $(\Z_2)^d$ center symmetry
    [such as $U(\Nc)$ theories with tensor representation matter fields,
    or $O(\Nc)$ pure gauge theories],
    the center symmetry only forces loops with odd winding numbers
    to have vanishing expectation values.
    Loops with even winding numbers can, and will, have non-zero
    expectations for finite values of $\Nc$.
    But these expectation values are $1/\Nc$ suppressed in the
    large $\Nc$ limit, provided the $(\Z_2)^d$ center symmetry
    is unbroken.
    This may be understood from the loop equations for observables
    with non-zero but even winding numbers.
    It is the splitting terms which act like source terms that
    generate non-zero expectation values for larger loops.
    In, for example, the equation for a winding number two observable,
    the source terms can involve products of winding number
    one loops, but cannot have products of only winding number
    zero loops.
    As long as the $(\Z_2)^d$ center symmetry is unbroken,
    the disconnected part of the expectation of a product of two
    winding number one loops will vanish, while the connected part
    is $1/\Nc$ suppressed.
    At $\Nc = \infty$
    there is, in essence, a topological conservation law which
    prevents non-zero winding number sectors from directly coupling
    to the zero winding number sector,
    even when the center symmetry, alone, would allow such coupling.
    }
Just as with adjoint representation matter, this equivalence
applies to single trace observables whose winding numbers,
and total momentum, are compatible with the spatial periodicity
in both volumes.

\subsection{Temperature dependence}

In our discussion of volume changing projections we have,
purely for ease of presentation,
focused on the case of mappings which
uniformly affect all dimensions.
However, all discussion of large $N$ spacetime volume independence applies
equally well to mappings which expand or contract different
dimensions independently.
The particular case of large $N$ equivalence under mappings which
change the size of just one periodic dimension is relevant for
understanding temperature dependence in large $N$ gauge theories
(since a non-zero temperature theory is merely one in which
a chosen Euclidean time direction has been compactified with
a period equal to the inverse temperature $\beta \equiv 1/T$).

Consider some large $N$ gauge theory at two different (inverse)
temperatures $\beta$ and $\beta'$.
As always, large $N$ equivalence applies only to
observables in appropriate neutral sectors,
and only if the symmetries defining these neutral sectors
are not spontaneously broken.
If $\beta = K \beta'$, for some integer $K$,
then the neutral sector in the lower temperature theory
consists of all single trace observables whose Matsubara frequencies
are integer multiples of $2\pi/\beta'$
(not just integer multiples of $2\pi/\beta$).
This, of course, is equivalent to the requirement that the observables
be invariant under Euclidean time translation by multiples of $\beta'$.
In the higher temperature theory, the neutral sector consists of all
single trace observables whose winding numbers around the thermal
circle are multiples of $K$.
The large $N$ limits of expectation values of corresponding neutral
observables coincide,
and their connected correlators satisfy the relation
(\ref {eq:connected-corr2}) (with $d = 1$),
{\em provided} the $\Z_K$ symmetry defining the neutral sector in the
higher temperature theory is not spontaneously broken.

In the context of thermal gauge theories
with only adjoint or tensor representation matter,
the realization of the center symmetry associated with the
thermal circle provides a sharp distinction between
a confining phase, in which the center symmetry is unbroken,
and a deconfined plasma phase, in which the center symmetry is
spontaneously broken.
Hence, the condition of unbroken center symmetry,
required for large $N$ volume independence,
is precisely the same as the requirement that the theory be in
a confining phase.

Therefore, within a confining phase, large $\Nc$ volume independence
implies that the $O(\Nc^2)$ part of the free energy is temperature
independent.
This is true, but trivial, in pure gauge theories, and theories
with only adjoint or tensor representation matter fields,
since temperature dependent contributions to
the free energy in the confining low temperature phase of
such theories only  arises from glueball-like bound states
whose spectral density is $O(\Nc^0)$.%
\footnote
    {
    Temperature independence of the free energy also applies
    to the $O(\Nc)$ part of the free energy if there are
    fundamental representation matter fields with $\Nf$ held
    fixed as $\Nc\to\infty$, but this temperature
    independence is again trivial since the thermal contribution
    of mesons is $O(\Nf^2)$ which, by assumption is $O(\Nc^0)$.
    However, if $\Nf$ scales with $\Nc$, so that their ratio is held fixed,
    then the free energy does have non-vanishing $O(\Nc^2)$
    temperature dependence and large $N$ temperature independence
    fails to hold.
    }

The implications of large $N$ temperature independence are non-trivial
for expectation values and connected correlators of neutral operators.
If $\beta$ and $\beta'$ are not rationally related,
then the neutral sector consists of single-trace observables
for which both the Matsubara frequency
and winding numbers around the thermal circle vanish.%
\footnote
    {
    If $\beta$ and $\beta'$ are rationally related,
    so that $\beta/M = \beta'/M'$ for coprime integers $M$ and $M'$,
    then the neutral sector is larger and consists of observables whose
    Matsubara frequencies are multiples of $2\pi/(\beta/M)$, and whose
    winding number in the temperature $\beta$ theory vanish modulo $M'$
    (or equivalently whose winding number in the $\beta'$ theory
    vanish modulo $M$).
    }
Large $N$ temperature independence implies, for example, that the
value of the chiral condensate in QCD (with $\Nf$ fixed),
or in $\None$ super-Yang-Mills, is temperature independent at
$\Nc=\infty$ in the confining phase.
The non-perturbative gluon condensate
$\langle F_{\mu\nu}^2 \rangle$
and the area law coefficient of spacelike Wilson loops must
likewise be temperature independent.
But large $N$ temperature independence does not apply to the
physical string tension, since this is determined by the
correlator of Wilson lines which wrap just once around the thermal circle,
and these operators do not lie within the relevant neutral sector.

\section{Eguchi-Kawai reduction for QCD}
\label{EKforQCD}

As reviewed above (and illustrated in Fig.~\ref{fig:phase}),
large $N_c$ volume independence for pure Yang-Mills theory
(in $d > 2$ dimensions)
is valid only above a critical size, $L> L_c(d)$.
Below this limit, spontaneous breaking of the $U(1)^d$ center symmetry
invalidates volume independence.
Adding fundamental representation matter fields
either has no effect on this breakdown of large $\Nc$ volume independence
(if $\Nf$ is held fixed), or completely destroys volume independence
(if $\Nf/\Nc$ is held fixed)
for large volumes.

The effect of adjoint or tensor representation matter fields on
volume independence is more interesting, since a single flavor
of such matter fields can have a significant impact on the dynamics
of the theory, even at $\Nc = \infty$.
If one considers QCD(AS) [or QCD(S)]
with one or more compactified dimensions,
then a simple perturbative analysis of the one-loop effective
potential for Wilson lines \cite{UY,Barbon},
which is reliable when the compactification radius is small compared to
$\Lambda_{\rm QCD}^{-1}$,
shows that the $(\Z_2)^d$ center symmetry is spontaneously broken.%
\footnote
    {
    The small volume phase also has unbroken chiral symmetry.
    See Ref.~\cite{UY} for further details.
    This analysis requires that $1 \le \Nf \le 5$.
    Adding more than five flavors destroys asymptotic freedom.
    }
(This is true regardless of whether one uses periodic or antiperiodic
boundary conditions for the fermions.)
But a confining phase with unbroken center symmetry
(and spontaneously broken chiral symmetry)
will surely exist in sufficiently large volume.
Therefore, large $\Nc$ volume independence in QCD(AS/S) fails below
some critical size, which must be of order $\Lambda_{\rm QCD}^{-1}$,
just as in pure Yang-Mills theory.

    The situation with adjoint representation fields is different.
Consider $U(\Nc)$ QCD(Adj) with one or more flavors of massless adjoint
Majorana fermions.
(Asymptotic freedom requires that $\Nf \le 5$.)
Let one dimension be compactified with a radius $L$ which is much smaller
than all other dimensions, as well as $\Lambda_{\rm QCD}^{-1}$,
and choose periodic boundary conditions for the adjoint fermions.
So the theory is
on $\R^3 \times S^1$, where the $S^1$ may be regarded as
a spatial circle.
The one-loop effective potential for the Wilson line around this
compactified direction is%
\footnote
    {
    A completely analogous, but more complicated, formula
    results if one compactifies two or more dimensions with radius $L$.
    The following discussion is unaffected.
    See Ref.~\cite{Barbon} for details.
    }
\begin{eqnarray}
    V_{\rm eff}^{\rm QCD(Adj)}(\Omega)
    &=&
    (\Nf {-} 1) \frac{2}{\pi^2 L^4}
    \sum_{n=1}^{\infty}\frac{1}{n^4} \> |\tr \, \Omega^n|^2
\nonumber
\\
    &=&
    \frac{\Nf {-} 1}{24\pi^2L^4}
    \biggl\{
    \frac {8\pi^4}{15} \, \Nc^2
    -
	\sum_{i, j=1}^{\Nc} [v_i -v_j]^2  ( [v_i -v_j]- 2\pi)^2
    \biggr\}
    \,.
\label{Eq:potential3}
\end{eqnarray}
Here $\{ e^{i v_j} \}$ denote the eigenvalues of the
(untraced) Wilson line $\Omega$,
and $[x] \equiv x \bmod 2\pi$ indicates quantities defined to lie
within the interval $[0, 2\pi)$.
Note that this effective potential is positive
for $\Nf >1$,
negative for $\Nf = 0$, and vanishes for $\Nf = 1$.
If $\Nf = 0$, then the potential generates
an attractive interaction between eigenvalues leading to
spontaneous breaking of the center symmetry and
a non-zero expectation value for the Wilson line.
This is just the pure Yang-Mills case discussed above,
in which center symmetry breaking invalidates volume independence
in small volumes.

If $\Nf > 1$, then the
potential generates a repulsive interaction between eigenvalues.
Consequently, the Wilson line eigenvalues distribute uniformly
around the unit circle, the Wilson line expectation value
$\langle \tr \, \Omega\rangle$ vanishes,
and the center symmetry is unbroken.%
\footnote
    {
    As noted above, we are viewing the $S^1$ as a spatial circle,
    and accordingly the relevant center symmetry is a spatial center symmetry.
    Its realization should not be interpreted as a confinement criterion,
    since it is temporal Wilson lines,
    not spatial Wilson lines, which are order parameters for
    confinement/deconfinement transitions.
    But if antiperiodic boundary conditions are chosen, then one may
    instead regard the $S^1$ as a temporal circle, in which case
    a small radius compactification is equivalent to the
    high temperature limit.
    For this choice, regardless of the number of fermions,
    the one-loop potential generates an attractive
    interaction between the Wilson line eigenvalues which causes
    spontaneous breaking of the (temporal) center symmetry,
    thereby signaling deconfinement and invalidating
    large $\Nc$ temperature independence.
    }

If $\Nf=1$, then the massless QCD(Adj) theory is just $\None$ supersymmetric
Yang-Mills theory.
For this case,
the one-loop potential identically vanishes.
Supersymmetry guarantees that this remains true at all loop orders.
However, there is an instanton induced non-perturbative superpotential
\cite{Davies:1999uw}.
The role of the non-perturbative potential in the $\Nf=1$ case is
essentially identical to the role of the
one-loop effective potential when $\Nf$ is greater than one.
It also provides eigenvalue repulsion
which leads to unbroken center symmetry
and vanishing Wilson line expectations, $\langle \tr\,\Omega\rangle=0$.
The instanton analysis is justified in sufficiently small volume,
but the simple dependence on the holomorphic coupling required by
supersymmetry guarantees that the result remains valid in arbitrary volume.

Consequently, QCD with $1 \le \Nf \le 5$ massless adjoint representation
fermions (and periodic boundary conditions) does not undergo a spatial
center symmetry breaking phase transition in small volume.%
\footnote
    {
    This is true for $SO(2\Nc)$ and $Sp(2\Nc)$ gauge groups
    as well as for $U(\Nc)$.
    }
If non-zero fermion masses are introduced,
this symmetry realization will be stable provided the
fermion masses are small compared to $\Lambda_{\rm QCD}$.%
\footnote
    {
    But as one raises the fermion mass,
    in sufficiently small volumes,
    there will be a phase transition to a center-symmetry broken
    phase at some critical mass of order $\Lambda_{\rm QCD}$,
    since the behavior of the theory must approach that of pure
    Yang-Mills theory when
    all fermion masses are large compared to the strong scale.
    }
Therefore, with at least one (and at most five)
light fermions,
QCD(Adj) is an asymptotically free, confining gauge theory which satisfies
large $\Nc$ volume independence all the way down to zero size.

To recap,
large $\Nc$ volume independence fails for sufficiently small volumes
in QCD(AS/S)
but remains valid in QCD(Adj)
(with periodic boundary conditions for fermions).
Despite this difference,
the validity of volume independence down to zero size in QCD(Adj)
has interesting implications for QCD(AS/S).
The key ingredient for this connection is the existence of
a large $\Nc$ equivalence \cite{ASV1,ASV2,UY} between QCD(AS/S) and QCD(Adj)
(for coinciding values of $\Nf$, `t Hooft coupling, fermion masses,
and boundary conditions)
provided that charge conjugation symmetry ($\C$) is not spontaneously broken
in QCD(AS/S).%
\footnote
    {
    This so-called ``orientifold equivalence'' is an example of a
    daughter-daughter orbifold equivalence;
    one can obtain either $U(\Nc)$ QCD(AS/S) or $U(\Nc)$ QCD(Adj)
    by applying  different orbifold projections to
    a common parent theory
    [namely, QCD(Adj) with either $SO(2\Nc)$ or $Sp(2\Nc)$ gauge group].
    See Ref.~\cite{UY} for further discussion.
    }
The appropriate neutral sector in QCD(AS/S) for this large $\Nc$ equivalence
consists of charge conjugation even single-trace bosonic operators.
Consequently, as long as charge conjugation symmetry is not
spontaneously broken in QCD(AS/S), expectation values and
connected correlators of corresponding $\C$-even single-trace
operators in QCD(AS/S) and QCD(Adj) will coincide at $\Nc = \infty$.

\begin{FIGURE}[ht]
    {
    \parbox[c]{\textwidth}
        {
        \begin{center}
	\psfrag{inf}{$\infty$}
        \includegraphics[width=0.9\textwidth]{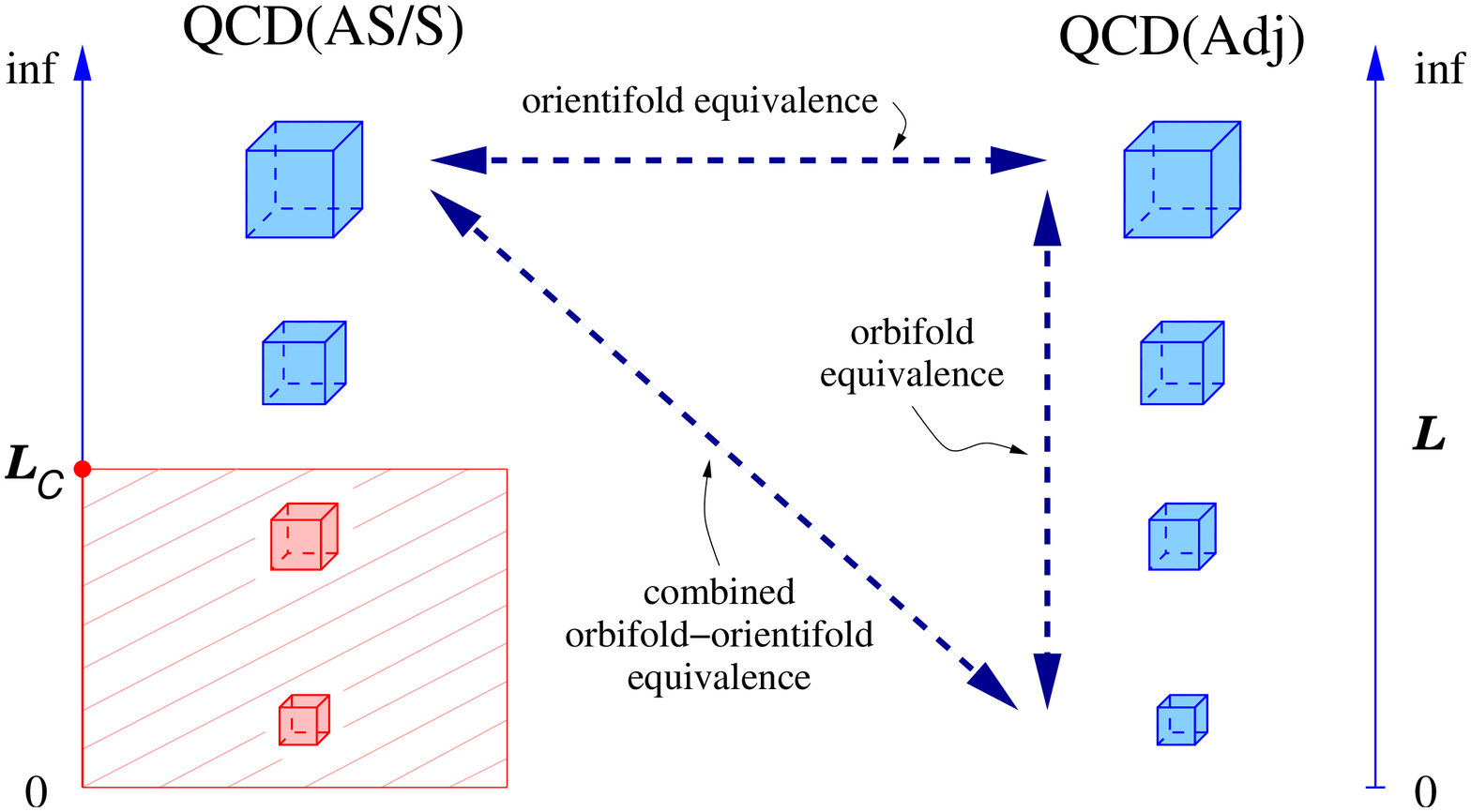}
        \caption
	{
	Large $\Nc$ equivalences relating QCD(AS/S) and QCD(Adj),
	as a function of the size $L$ of the periodic volume
	in which the theories are defined.
	Provided periodic boundary conditions are used for fermions,
	volume independence holds in QCD(Adj) for all $L$.
	But in QCD(AS/S), volume independence fails below a critical
	size, $L<L_\C$, (shaded region)
	due to spontaneous breaking of center symmetry.
	This prevents reduction all the way down to a single-site matrix model
	for QCD(AS/S).
	Large $\Nc$ orientifold equivalence holds between QCD(Adj)
	and QCD(AS/S) as long as charge conjugation symmetry is unbroken
	in QCD(AS/S).
	This should be true in sufficiently large volumes,
	but fails when $L < L_\C$.
	The combination of volume changing orbifold projections in QCD(Adj)
	along with the orientifold equivalence in large volume provides
	a useful equivalence between small volume QCD(Adj)
	and large volume QCD(AS/S).
	In particular, a single-site matrix model of QCD(Adj)
	will reproduce properties of infinite volume QCD(AS/S).
	}
        \label {fig:volind}
        \end{center}
        }
    }
\end{FIGURE}

In sufficiently small volume, charge conjugation symmetry {\em is\/}
spontaneously broken in QCD(AS/S) if one uses periodic boundary
conditions for fermions.
In this case,
the spontaneous breaking of the $(\Z_2)^d$ center symmetry
mentioned above is associated with the development of an imaginary
expectation for the Wilson line, which indicates spontaneous breaking
of charge conjugation (as well as parity and time reversal) symmetry,
in addition to the broken spatial center symmetry \cite{UY}.
This symmetry breaking is sensitive to the choice of boundary conditions for
the fermions and is evidently a finite size effect.
There is no known reason to expect spontaneous breaking of charge
conjugation symmetry in sufficiently large volumes.
For the case of $\Nc = 3$ and $\Nf = 4$,
recent lattice simulations \cite{DeGrand:2006qb}
have clearly seen the presence of a small volume phase in QCD(AS)
with spontaneously broken charge conjugation symmetry,
together with clear evidence for a phase transition consistent with
restoration of charge conjugation symmetry
at a critical size very near the inverse strong scale.
[The chiral symmetry realization also changes at this transition,
indicating the presence of a chirally symmetric phase of QCD,
at zero (or low) temperature,
in sufficiently small volume.]
Therefore, it seems reasonable to believe that QCD(AS/S) does
have unbroken charge conjugation symmetry in large volume.%
\footnote
    {
    However, there is no rigorous proof of this.
    The Vafa-Witten theorem on $\R^4$ \cite{Vafa-Witten} demonstrates
    unbroken parity, but does not determine the
    charge conjugation symmetry realization.
    In fact,
    the theorem only rules out cases of spontaneous breaking of parity with
    a local order parameter.
    In a compactified theory, breaking of parity
    and center symmetry can be entangled so that
    order parameters must involve topologically non-trivial Wilson lines.
    Spontaneous symmetry breaking in this case is not ruled out by
    the Vafa-Witten theorem [and occurs in just this manner in QCD(AS/S)].
    }

Consequently, the large $\Nc$ equivalence between QCD(AS/S)
and QCD(Adj) should be valid above a critical size,
while large $\Nc$ volume independence in QCD(Adj) is valid for all sizes.
Hence,
even though volume independence fails in QCD(AS/S) in sufficiently
small volume,
{\em QCD(AS/S) in large volume has a large $\Nc$ equivalence
relating it to QCD(Adj) in arbitrarily small volumes.}
This is depicted schematically in Fig.~\ref{fig:volind}.

This equivalence applies to observables in the intersection of the
neutral sectors for the orientifold equivalence and volume
changing orbifold projections.
Explicitly, these are $\C$-even single-trace observables
whose total momenta are compatible with the small volume.
In particular, a single-site version of QCD(Adj),
which is just the EK matrix model (\ref{eq:EKaction})
augmented with adjoint representation Grassmann variables,
will reproduce the leading large $\Nc$ behavior of all
expectation values, and zero-momentum connected correlators, of
arbitrary $\C$-even single-trace observables in infinite volume QCD(AS) ---
which is a natural generalization of real QCD to large $\Nc$.
%
%
This is a version of Eguchi-Kawai reduction of large $\Nc$ QCD
which works all the way down to zero size.

\section{Theory space volume independence}
\label{sec:quiver}

The spacetime volume independence discussed in the previous section
has a natural counterpart involving large $N$ equivalences
among quiver gauge theories.
It is convenient to represent the gauge group and matter content
of such theories by a ``theory space graph'' (or ``quiver diagram''),
as illustrated in Fig.~\ref{fig:theoryspace}.
Each node represents a simple gauge group factor, while each bond
represents a matter field.
A directed bond which connects two nodes represents a
matter field transforming under the fundamental representation
of the gauge group factor where it starts, and the
anti-fundamental representation of the gauge group factor
where it ends.
A bond which begins and ends at the same node represents a
matter field transforming under the adjoint representation
of the corresponding gauge group factor.
Note that
in any quiver theory containing only adjoint or bifundamental
matter fields, gauge invariant single-trace operators
must correspond to closed loops in theory space.
(For more detailed discussion see,
for example, Refs.~\cite{Rothstein:2001tu, Arkani-Hamed:2001ed}.)

\begin{FIGURE}[ht]
    {
    \parbox[c]{\textwidth}
        {
        \begin{center}
	\vspace*{-1em}
        \includegraphics[width=0.8\textwidth]{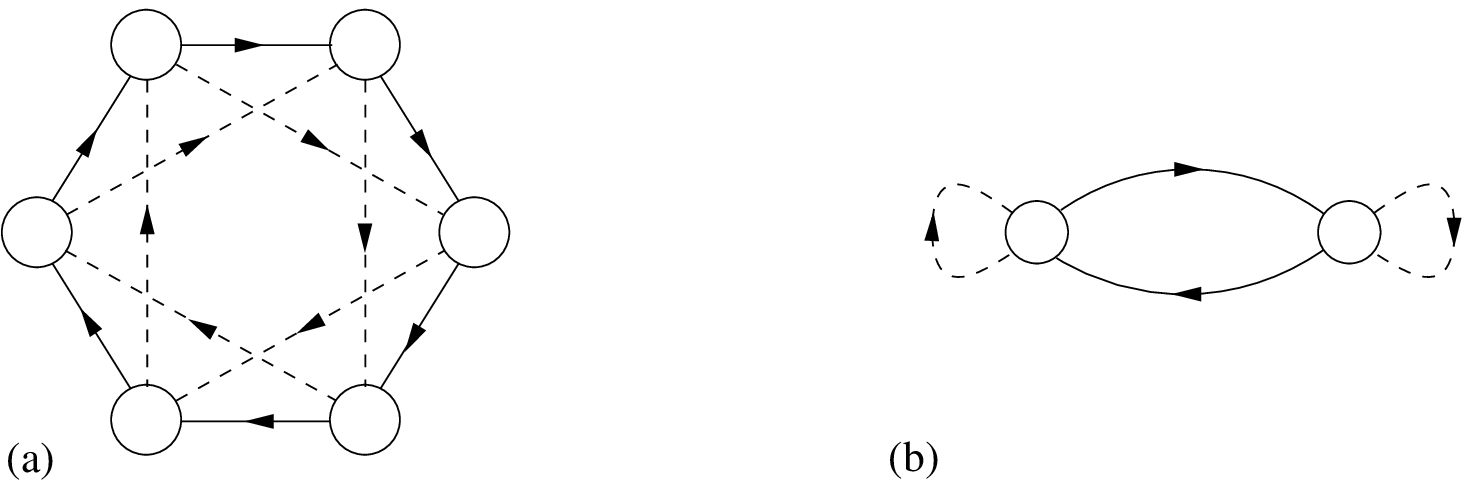}
	\vspace*{-1em}
        \caption
	{
	Examples of theory space graphs.
	Graph $(a)$ shows the case of a $U(N)^6$ gauge theory
	with $12$ bifundamental representation matter fields;
	one set, represented by solid bonds,
	transforms under ``nearest neighbor'' group factors
	while the other set, represented by dashed bonds,
	transforms under next-nearest neighbor group factors.
	This graph (and the underlying theory) has an obvious
	$\Z_6$ symmetry corresponding to cyclic permutations of
	the different group factors.
	Graph $(b)$ shows the case of a $U(N)^2$ theory
	which may be obtained from theory $(a)$
	by applying a projection based on the $\Z_3$ subgroup of the
	theory space symmetry group.
	}
        \label {fig:theoryspace}
        \end{center}
        }
    }
\end{FIGURE}

\subsection{Volume reducing projections}

If the theory has a global symmetry which interchanges
or permutes equivalent gauge group factors, then this symmetry
will be a geometric invariance of the theory space graph.
A simple example is a parent theory with
a $U(N)^{L^d}$ gauge group and $\Nf \, L^d$ bifundamental matter fields
coupling the different gauge group factors in such a way that the
resulting theory has a $(\Z_L)^d$ symmetry which cyclically
permutes gauge group factors.
Fig.~\ref{fig:theoryspace}(a) illustrates a case of $L=6$, $d = 1$,
and $\Nf = 2$.

Let $\cal T$ be some chosen subgroup of the global symmetries
involving theory space permutations.
One may define a projection which eliminates degrees of freedom
that are not invariant under the subgroup $\cal T$.
This is equivalent to identifying nodes (and bonds) of the theory
space graph which are related by transformations in $\cal T$.
The result will be a daughter theory with a theory space which
is smaller than that of the parent,
as illustrated in Fig.~\ref{fig:theoryspace}.

In the example of a parent theory with $(\Z_L)^d$ theory space symmetry,
consider the case of $\mathcal T = (\Z_K)^d$, where $L$ is divisible
by $K$.
The projection will yield a daughter theory with a smaller
$(\Z_{L'})^d$ theory space, with $L' \equiv L/K$.
The discussion in section \ref{sec:volume-reduction},
concerning projections based on a subgroup of the spacetime
translation symmetry group,
may be reapplied almost verbatim to this class of
projections which reduce the size of theory space.

The neutral sector of the parent theory consists of
gauge invariant single-trace operators which are
invariant under the projection subgroup $\cal T$.
The neutral sector of the daughter theory consists of
gauge invariant single-trace operators whose winding numbers
around the periodic cycles in theory space all vanish modulo $K$.
The projection defines a one-to-one mapping between
these two neutral sectors.

Once again, the action (or free energy) of the parent
and daughter theories are related by an overall factor of the
size of the projection subgroup,
\begin{equation}
    S_{\rm parent} \longrightarrow K^d \, S_{\rm daughter} \,.
\label{eq:Smap3}
\end{equation}

\subsection{Volume enlarging projections}

One may also define projections which increase the size of
theory space, in a manner completely analogous to
the previous discussion of spacetime volume enlarging projections.
The only difference between the two cases is that for quiver theories,
it is the bifundamental matter fields which transform non-trivially
under two different gauge group factors, while for a lattice gauge
theory it is the link variables which transform non-trivially under
gauge group factors located at two different sites.

The most commonly discussed examples of orbifold projections
map a $U(\Nc)$ theory with a simple gauge group into a quiver theory.
(See, for example, Refs.~\cite{Rothstein:2001tu,Arkani-Hamed:2001ed,KUY1}.)
These cases correspond precisely to the volume enlarging projections of the
single-site Eguchi-Kawai model discussed above.
A $(\Z_K)^d$ projection, for example, corresponds to constraints of the form
\begin{equation}
    A_\mu = \gamma_\nu\, A_\mu \, \gamma_\nu^\dagger \,,
    \qquad \nu = 1,\cdots,d \,,
\end{equation}
for the gauge fields (viewed as $\Nc \times \Nc$ matrices),
together with
\begin{equation}
    \Phi_a = \gamma_\nu\, \Phi_a\, \gamma_\nu^\dagger \>
    e^{2\pi i r_\nu[\Phi_a]/K} \,,
    \qquad \nu = 1,\cdots,d \,,
\end{equation}
for matter fields.
Here $\Nc$ must be divisible by $K^d$ and
$\{ \gamma_\nu \}$ are the same mutually commuting matrices
which appeared in Eq.~(\ref{eq:EKconstraint}) and are defined explicitly in
Eq.~(\ref{eq:gamma-matrices}).

Each matter field $\Phi_a$ is assigned a $d$-dimensional integer-valued
charge vector with components $ r_\nu[\Phi_a] $.
Different choices for the
charge vector assigned to each field
correspond to different embeddings of $(\Z_K)^d$ into the product
of the gauge and flavor symmetry groups, and produce
daughter theories with differing connectivities in their theory space graphs.
If the charge vector assigned to a particular field vanishes,
then field becomes an adjoint representation field in the daughter
theory.
If the charge is non-zero, then the field becomes a bifundamental,
with the value of the charge vector determining the connectivity
of the corresponding bonds in the theory space graph.

The neutral sector in the parent $U(\Nc)$ theory
consists of gauge invariant single-trace operators for which
the sum of the charge vectors of all matter field insertions vanish.
The neutral sector in the daughter $[U(N)]^{K^d}$ theory
consists of gauge invariant single-trace operators which are
invariant under $(\Z_K)^d$ cyclic permutations of the gauge group
factors (corresponding to translations in the periodic theory space graph).
Once again, the projection defines a one-to-one mapping between
these two neutral sectors, with the appropriate relation between
the actions of the two theories involving a rescaling by the size
of the projection,
\begin{equation}
    S_{\rm parent} \longrightarrow K^d \, S_{\rm daughter} \,.
\end{equation}

Instead of starting from a parent theory with a simple gauge group,
one may define analogous theory-space enlarging projections
starting from any quiver theory.
This is done analogously to the volume expansion procedure
of the lattice gauge theory, as discussed in the Appendix.

\subsection{Large $N$ equivalence}

The entire discussion of large $N$ equivalence in section
\ref{sec:largeN} applies equally well to the above examples
of theories related by $(\Z_K)^d$ theory-space enlarging
or reducing projections.
The $N = \infty$ dynamics within corresponding neutral sectors
coincides.
If the ground (or equilibrium) states of both theories lie within
their respective neutral sectors, then the large $N$ equivalence
implies coinciding expectation values of corresponding neutral
single-trace operators, and connected correlators of such operators
related by Eqs.~(\ref{eq:connected-corr2}) and (\ref{eq:connected-corr3})
(with $\langle \cdots \rangle^{N,L}$ now denoting an expectation in a
$[U(N)]^{L^d}$ quiver theory with
a $(\Z_L)^d$ invariant theory space).

The condition that the ground (or equilibrium) state of a theory
lie within its neutral sector is precisely the requirement that
the symmetries defining the neutral sector not be spontaneously broken.
The symmetry realization will, inevitably, depend on the specific
dynamics of a theory.
This includes not just the field content of the theory
plus masses and couplings of matter fields,
but also, for example, the temperature
and the spatial volume  in which the theory is defined
(as well as the associated boundary conditions).

The most well-studied example of large $N$ equivalence,
the $\Z_2$ projection of $\None$ supersymmetric
$U(2N)$ Yang-Mills theory, yielding a non-supersymmetric
$U(N) \times U(N)$ theory with a bifundamental fermion,
turns out to be remarkably similar to the case of spacetime
volume independence in pure Yang-Mills theories discussed above.
For this $\Z_2$ projection of super-Yang-Mills,
a useful large $N$ equivalence
requires that the $\Z_2$ symmetry exchanging gauge group
factors in the non-supersymmetric quiver theory be unbroken.
However, this symmetry is known to be spontaneously broken when
the theory is compactified (with periodic boundary conditions)
on $\R^3 \times S^1$,
with the radius of the $S^1$ small compared to the
inverse confinement scale, $R \ll \Lambda^{-1}$ \cite{Tong}.
In large volume,
there is no evidence that this $\Z_2$ symmetry is broken,
nor is there any proof (or solid evidence) that it is unbroken
\cite{KUY3}.
At sufficiently high temperature, $T \gg \Lambda$, it is clear that
this $\Z_2$ symmetry is unbroken \cite{UY}.
Hence, large $N$ equivalence to $\None$ super-Yang-Mills is valid
at sufficiently high temperatures, and may be valid at low
temperatures in sufficiently large volume.

\section{Discussion}

This paper has attempted to highlight the direct connection between
volume (and temperature) independence in large $N$ gauge theories and
large $N$ equivalences involving orbifold projections
and quiver gauge theories.
It is curious that nearly all discussion in the literature
concerning large $N$ orbifold equivalence has focused on
projections which increase the size of theory space,
while nearly all previous discussion regarding
large $N$ spacetime volume dependence has focused on
volume reducing mappings.
As we have emphasized, one can easily define projections
which either increase or decrease spacetime volume,
or ``theory space'' volume.
This unified view makes clear the common origin of
the symmetry realization conditions which are necessary (and sufficient)
for useful large $N$ equivalence in all these examples.

Combining the large $\Nc$ volume independence of QCD with
light adjoint fermions, valid down to arbitrarily small size
(when periodic boundary conditions are used for fermions),
with the large $\Nc$ ``orientifold equivalence'' relating
QCD with adjoint and antisymmetric tensor representation fermion,
which is valid in sufficiently large volumes,
produces a large $\Nc$ equivalence between a single site model
with adjoint fermions, and a natural large $\Nc$ generalization
of real QCD.
Although this equivalence only applies to charge conjugation even
observables, it should nevertheless have practical utility
for investigations of large $\Nc$ QCD.
In particular, this form of a large $\Nc$ reduced model
is applicable to both expectation values and
suitable connected correlation functions
(in contrast to quenched and twisted reduced models).
Although simulations with dynamical fermions are always more challenging
than pure gauge simulations,
simulating a single site model with adjoint representation fermions
should be relatively straightforward.

Many generalizations of the specific examples discussed above
are possible.
One natural generalization involves
consideration of $O(\Nc)$ or $USp(\Nc)$ gauge theories instead
of $U(\Nc)$.
We discuss the interconnections between such theories in
a companion paper \cite{KUY5}.
In quiver gauge theories,
one may also consider theory-space enlarging projections based on
non-Abelian, or non-freely acting, symmetry groups \cite{Feng:2000af}.
As noted in the Introduction, for general projections
based on arbitrary discrete symmetry groups,
it is not clearly established when such projections lead to large $N$
equivalence, and when they do not.
But every case of large $N$ equivalence that we are aware of
is consistent with the conjecture
that pairs of theories connected by
``invertible'' projections imply a large $\Nc$ equivalence between
the two theories.
(This conjecture is stated more formally in the Introduction.)
Future work will hopefully shed more light on this issue
of a global understanding of equivalences between different
large $N$ gauge theories.

\begin{acknowledgments}

We thank H.~Neuberger for useful comments on the subject of reduced models.
This work was supported, in part,
by the U.S. Department of Energy under Grant Nos.~%
DE-AC02-76SF00515,
DE-FG02-91ER40676,
and DE-FG02-96ER40956,
and the National Science Foundation under Grant No.~PHY05-51164.

\end{acknowledgments}

\appendix

\section {More on volume expansion as an orbifold projection}

The main text discussed volume-expanding orbifold projections
starting from single-site models.
In this appendix we discuss the more general case of projections which
will map a parent theory defined on a periodic lattice of any size
to a daughter theory defined on a lattice which is larger by some
(integer) factor $K > 1$.
The lattice in question can be either the real space lattice
on which the theory is defined, or the theory space of a quiver
gauge theory.
For simplicity of presentation,
we will focus on pure gauge theories defined on a real space lattice,
but what follows equally applies to volume expansion in theory space
of quiver gauge theories.

Start with a parent lattice gauge theory
on lattice $\Lambda' = ({\mathbb Z}_{L'})^d$
with gauge group $U(\Nc)$, where $\Nc{=}M^d N$
for some integer $M > 1$.
The global symmetry group of the theory is
$U(M^d N){\times}U(1)^d $,
where the first factor is the global part of the gauge symmetry group and
the $U(1)^d$ is the center symmetry which phase rotates all link variables
in a given direction, $U_\mu[\n']\to \exp(i\theta_\mu) U_\mu[\n']$.
Choose a cyclic subgroup $(\Z_M)^d\subset U(M^d N)\times U(1)^d$
of the global symmetry group
and set all $(\Z_M)^d$-noninvariant fields of the theory to zero.
The appropriate embedding is the one for which
the net effect of the projection is the imposition of the constraints
\begin{equation}
    U_\mu[\n'] =
      \omega^{\delta_{\mu\nu}}\; \gamma_\nu\,U_\mu[\n']\,\gamma_\nu^{-1}\,,
      \qquad \mu,\nu = 1, \cdots, d \,.
   \label{eq:orbifold-constraint-EK-U}
\end{equation}
Here, the matrices $\gamma_\nu\in U(M^d N)$ generate a
$M^d N$-dimensional representation of $(\Z_M)^d$
and are chosen as
\begin{equation}
   \gamma_\nu
   =
   \underbrace{1_{M} \times\dots}_{\nu-1}\times\,\Omega
   \times
   \underbrace{1_{M} \times\dots}_{d-\nu}
   \times \,1_N
\label{eq:gamma-matrices}
\end{equation}
where $1_N$ and $1_M$ are $N{\times}N$ and $M{\times}M$
unit matrices, respectively, and
\begin{equation}
  \Omega \equiv {\rm diag}(1,\omega,\omega^2,\dots,\omega^{M-1})
\end{equation}
with $\omega=e^{2\pi i/M}$.
The phase factor $\omega^{\delta_{\mu\nu}}$
in Eq.~(\ref{eq:orbifold-constraint-EK-U})
reflects a transformation under the
$U(1)^d$ part of the symmetry group.
This particular choice of the charge
for the link fields guarantees that the daughter theory will retain
nearest neighbor interactions on an enlarged lattice.
[Adjoint representation matter fields in the theory
would be neutral under the $U(1)^d$ symmetry, and
hence would have no additional phase factor in their orbifold constraints.]

In order to find the form of link matrices $U_\mu[\n']$ which satisfy the
orbifold constraint (\ref{eq:orbifold-constraint-EK-U}),
it is helpful to introduce $M^d N{\times}M^d N$
``translation matrices'' $T_\mu$ defined as
\begin{equation}
   T_\mu =
   \underbrace{1_M\times\dots}_{\mu-1}\times S\times
   \underbrace{1_M\times\dots}_{d-\mu}\times 1_N \ ,
\label{eq:translation-matrix}
\end{equation}
where $S$ is a $M{\times}M$ ``shift matrix'' whose entries
(defined modulo $M$) are $S_{ab}=\delta_{a,b-1}$.
Different translation matrices commute
with each other, $[T_\mu,T_\nu]=0$.
Using the definition of the matrices $\gamma_\nu$, one finds
that the $T_\mu$ satisfy the same orbifold constraint
(\ref{eq:orbifold-constraint-EK-U}), namely
\begin{equation}
  T_\mu = \omega^{\delta_{\mu\nu}}\, \gamma_\nu T_\mu \gamma_\nu^{-1}\ .
\end{equation}
Thus if one defines $U_\mu[\n']=\widetilde U_\mu[\n'] \, T_\mu$,
then the redefined matrices $\widetilde U_\mu[\n']$ satisfy
a simpler ``neutral'' constraint
\begin{equation}
    \widetilde U_\mu[\n'] =
    \gamma_\nu \, \widetilde U_\mu[\n'] \, \gamma_\nu^{-1}\ ,
\label{eq:orbifold-constraint-EK-neutral}
\end{equation}
without any additional phase factor.
In other words,
all redefined link matrices $\widetilde U_\mu[\n']$
must commute with the projection matrices $\{ \gamma_\nu \}$.
To find the most general solution of these constraints,
it is helpful to note that every $M{\times}M$ matrix which commutes
with $\Omega$ must be diagonal, and hence may be expressed as an
order $M$ polynomial in $\Omega$.
Consequently, any link variable $\widetilde U_\mu[\n']$ that
commutes with all $\gamma_\nu$'s must have the form
\begin{equation}
   \widetilde U_\mu[\n'] = \sum_{\bf p} \>
   \Omega^{p_1}\times\dots\times\Omega^{p_d}\times \widetilde V_\mu[\n',{\bf p}]
\end{equation}
where ${\bf p}=(p_1,\dots,p_d)$ is a vector whose
components are integers running from $0$ and $M{-}1$ (defined modulo $M$).
This is a mixed representation in which ${\bf n}$ is a real space label but
${\bf p}$  is a momentum space vector lying in a Brillouin zone.
It is far more convenient to express  $\widetilde U_\mu[\n']$ in a
``position basis'' via the discrete Fourier transform,
\begin{equation}
   \widetilde V_\mu[\n',{\bf p}] = \frac{1}{M^d}
   \sum_{\bf m} \> V_\mu[\n',{\bf m}]\> \bar\omega^{{\bf p}\cdot{\bf m}}
\end{equation}
where $\bar\omega=e^{-2\pi i/M}$ is complex conjugate of $\omega$,
and ${\bf m}$ is a $d$-dimensional vector whose components
are integers ranging from $0$ to $M{-}1$ (modulo $M$).
Then a matrix $\widetilde U_\mu[\n']$ that solves the neutral constraint
(\ref{eq:orbifold-constraint-EK-neutral}) has the form
\begin{equation}
   \widetilde U_\mu[\n'] =
   \sum_{{\bf m}} \> \Delta_{\bf m}\times V_\mu[\n',{\bf m}]\,,
\label{eq:constraint-solution-EK-U}
\end{equation}
where the ``basis matrices'' $\Delta_\m$ are given by
\begin{equation}
    \Delta_{\bf m} \equiv \frac 1 {M^d} \sum_{\bf p} \>
 \Omega^{p_1}\times\dots\times\Omega^{p_d}\>
 \bar\omega^{{\bf p}\cdot{\bf m}} \,,
\end{equation}
and each $V_\mu[\n',\m]$ is an arbitrary $N \times N$ unitary matrix.
The basis matrices are mutually orthogonal projectors and satisfy
\begin{equation}
   \tr\,\Delta_{\bf m} = 1\,, \qquad
   \Delta_{\bf m} \, \Delta_{\bf m'} =
   \delta_{\bf m m'} \, \Delta_{\bf m}\,, \qquad
   \sum_\m \Delta_\m = 1_{M^d} \,.
\label{eq:Deltas-properties}
\end{equation}
If the matrix $\widetilde U_\mu[\n']$ is viewed
as a collection of $M^d{\times}M^d$ blocks
(each of which is $N{\times}N$)
then only $M^d$ of those blocks (labeled by ${\bf m}$)
survive the constraint.

With matrices $\widetilde U_\mu[\n']$ of the form
(\ref{eq:constraint-solution-EK-U}),
it is not difficult to show that
the Wilson action of the $U(\Nc)$ lattice gauge theory on $\Lambda'$
becomes the Wilson action of a $U(N)$ lattice gauge theory on a larger lattice.
This is not immediately obvious because
each site of the new lattice has been labeled by two indices,
$\n'$ and $\m$.

The size of the new lattice depends on whether $M$ and $L'$ have
any common divisors.
If $M$ and $L'$ are coprime [so that $\gcd(M,L')=1$],
then the new lattice $\Lambda = \Z_L^d$ with $L = M L'$.
This reflects the isomorphism
$\Z_M \times \Z_{L'} \sim \Z_{ML'}$,
as illustrated in figure \ref{fig:project}a.
Under the isomorphism, the link variable $V_\mu[\n',\m]$
connects $[\n',\m]$ to $[\n'{+}{\bf e}_\mu,\m{+}{\bf e}_\mu]$,
which (by definition) is a nearest neighbor on the new lattice.

\FIGURE[t]
{
\begin{minipage}[t]{\textwidth}
\begin{center}
\psfrag{nprime}{$n'$}
\psfrag{m}{$m$}
\includegraphics[width=0.9\textwidth]{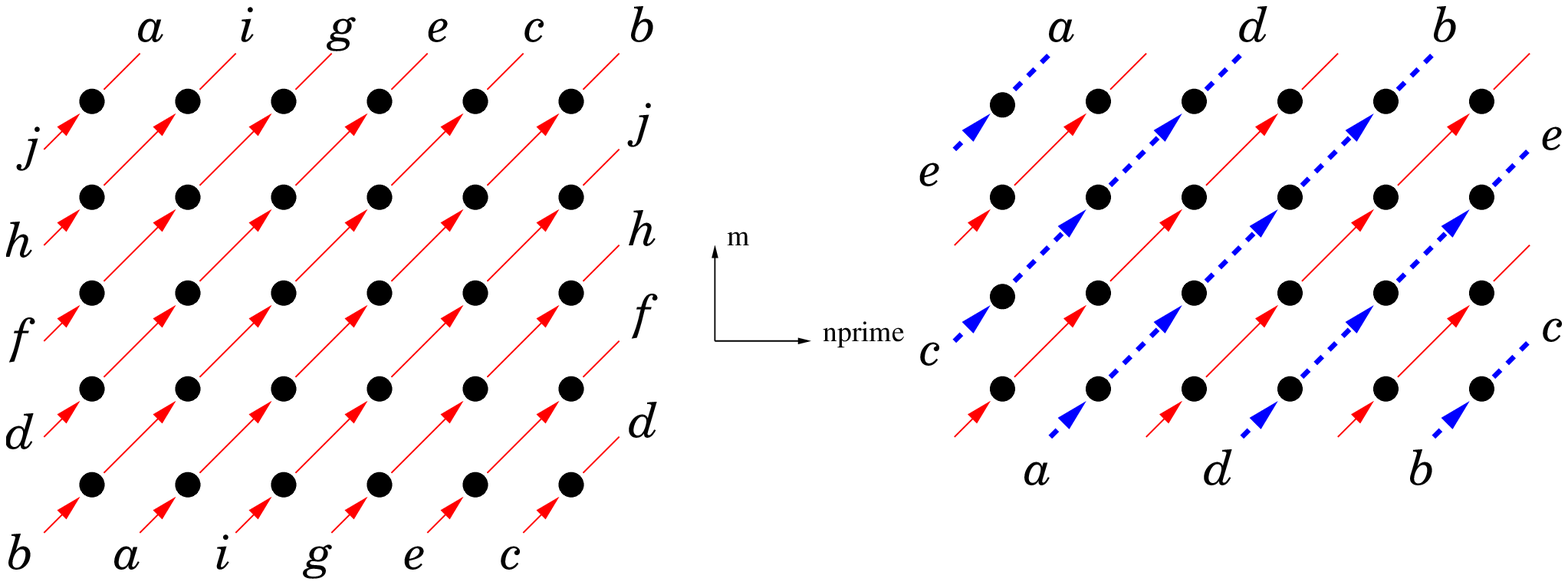}
\end{center}
\end{minipage}
\caption{
  Examples of one-dimensional daughter theory lattices generated by
  $\Z_5$ (left figure) or $\Z_4$ (right figure) projections acting on
  a parent theory defined on a six site lattice
  (with ${\mathbb Z}_6$ translation symmetry).
  Each site in the daughter lattice has been
  labeled by two integers $n'$ and $m$
  (even though these are one-dimensional lattices).
  Nearest-neighbor relations are defined by incrementing
  both $n'$ and $m$ by one.
  The $\Z_5$ projection
  yields a 30 site lattice with ${\mathbb Z}_{30}$ translation symmetry.
  The path formed by the nearest-neighbor steps depicted as red arrows
  forms a single cycle around the daughter lattice.
  (Labels at the edge of the lattice show periodic identifications.)
  In contrast, a $\Z_4$ projection acting on the same parent theory
  yields two decoupled copies of a twelve site lattice with
  ${\mathbb Z}_{12}$ translation symmetry.
  In this case, no sequence of nearest-neighbor steps
  can connect the two copies of the resulting daughter lattice.
} 
\label{fig:project}
} 

When $M$ and $L'$ are not coprime,
$\Z_{M} \times \Z_{L'}$ is isomorphic to $\Z_{L'M/Q}\times \Z_Q$,
where $Q$ is the greatest common divisor of $M$ and $L'$.
In this case, applying the orbifold projection
to the action of the parent theory on lattice $\Lambda'$
yields $Q^d$ decoupled copies of a daughter theory defined
on the lattice $\Lambda = \Z_L^d$ with $L \equiv ML'/Q$.
Figure  \ref{fig:project}b
shows an example where $d=1$, $L'=6$, and $M=4$.
For this case,
the outcome of the orbifold projection is two decoupled copies of the
larger volume theory on a 12 site lattice.
It is not difficult to show that neutral operators
in the parent theory (on lattice $\Lambda' = \Z_{L'}^d$)
map to neutral operators in the daughter theory
(on lattice $\Lambda = \Z_{ML'/Q}^d$).

A $(\Z_M)^d$ projection does not produce volume enlargement by a
factor of $M^d$  unless $\gcd(M,L')=1$.
However, by suitably choosing the projection
one can expand the volume by any desired integer factor.
To expand the volume $(L')^d$ by a factor of $K^d$, one starts with
a gauge theory whose number of colors is $\Nc=M^d N$,
and projects by $(\Z_M)^d$, where $M$ is chosen to satisfy
\begin{equation}
    L'M/\gcd(L',M)=L'K \,.
\end{equation}
This equation for $M$ always has at least one solution, $M=L'K$.
The projection yields a daughter theory on a lattice of the
desired size $(L'K)^d$
[or rather $\gcd(L',M)^d$ copies thereof].

\sloppy

\end {document}